\documentclass[a4paper, amsfonts, amssymb, amsmath, reprint, 
showkeys, nofootinbib, twoside,twocolumn]{revtex4}
\usepackage[english]{babel}
\usepackage[utf8]{inputenc}
\usepackage{multirow}
\usepackage{hhline}
\usepackage{tikz}
\usepackage{makecell}
\usepackage{comment}

\usepackage{caption}
\newcommand{\mc}[3]{\multicolumn{#1}{#2}{#3}}
\definecolor{tcA}{rgb}{0,0,0}
\usepackage{array}
\newcolumntype{P}[1]{>{\centering\arraybackslash}p{#1}}

\usepackage{soul}
\usepackage{graphicx}

\usepackage[titletoc]{appendix}

\usepackage[pdftex, pdftitle={Article}, pdfauthor={Author}]{hyperref} 
\begin{document}
\title{
Enlighting the transverse structure of the proton via double parton scattering in photon-induced interactions}

 \author{{ F. A. Ceccopieri}}
        \affiliation{ { Université Paris-Saclay, CNRS, IJCLab, 91405, Orsay, France \\ IFPA, Univèrsité de Liège, B4000, Liége, Belgium}}

\author{M. Rinaldi}
    \email[Correspondence email address: ]{matteo.rinaldi@pg.infn.it}
    \affiliation{Dipartimento di Fisica e Geologia. Università degli studi di Perugia. INFN section of Perugia. Via A. Pascoli, Perugia, 06123, Italy.}

\date{\today} 

\begin{abstract}

In the present paper 
we address double parton scattering (DPS) in 
quasi-real photon-proton interactions.
By using electromagnetic and hadronic 
models of the photon light cone wave functions,
we compute the so-called effective cross section, $\sigma_{eff}^{\gamma p}$
which allows us to calculate the DPS contribution to these processes under  dedicated assumptions. In particular, for the four-jet photoproduction in HERA kinematics we found a sizeable DPS contribution.
We show that 
if the photon virtuality $Q^2$ could be measured and thus the dependence of $\sigma_{eff}^{\gamma p}$ on such a parameter exposed, information on the transverse distance between partons active in proton could be extracted. To this aim, 
 we set lower limits on the integrated luminosity needed to observe such an effect which would allow the extraction of novel information on the proton structure.


\end{abstract}


\maketitle

\section{Introduction}

There are increasing experimental evidences that multiple parton interactions (MPI) may occur within the same hadronic collision, as a result of the composite and extended nature of the colliding hadrons. 
After preliminary investigations 
\cite{Goebel:1979mi,Humpert:1983pw,Mekhfi:1983az,Mekhfi:1985dv,Humpert:1984ay,Mangano:1988sq},
the inclusion of MPI
has been proven to be  required to obtain a proper description
of the 
multiplicity and topology of the hadronic final state 
{ of $pp$ collisions at collider energies}
\cite{1a,Sjostrand:1986ep}.
{ Whereas MPI are characterized by soft and semi-hard components, 
in the present letter we will focus on double parton scattering (DPS), 
in which both scatterings involve a large momentum transfer,
{ of the order of few GeV},
so that short distance cross sections are perturbatively calculable.}
DPS could unveil parton correlations in the hadron 
structure not accessible in single parton scattering (SPS). Such correlations are encoded in novel distributions, i.e. double Parton Distribution Functions (dPDFs) which appear in the DPS cross section. The latter are interpreted as the  number densities of a parton pair  with a given  transverse distance $b_\perp$  and carrying  longitudinal momentum fractions 
($ x_1,x_2$) of the parent hadron \cite{4a,Diehl:2011tt,5a,noi1,blok_1,Blok1,Blok2}.
Despite the ongoing theoretical efforts to investigate dPDFs  \cite{gauntevo,man,noij2,noij1,Diehl:2020xyg,Diehl:2018kgr}, the structure of the DPS cross section and its factorization properties \cite{Diehl:2017kgu,Diehl:2015bca,Gaunt:2014ska,Diehl:2019rdh,Diehl:2018wfy,Ryskin:2011kk,Ryskin:2012qx}, 
 rather limited { knowledge} has been accumulated so far and DPS 
measurements have  provided information { mainly on} $\sigma_{eff}$ 
 in $pp$ collisions \cite{Calucci:1999yz} 
 { and recently in $pA$ collisions \cite{Aaij:2020smi}}.
This dimensionful parameter controls the magnitude of DPS contribution under the simplifying assumptions of two uncorrelated hard scatterings and full factorisation of dPDFs in terms of ordinary PDFs and model-dependent distribution in transverse position space.
It has been shown in Refs. \cite{rapid,jhepc} that the knowledge of $\sigma_{eff}$ can provide 
information on the proton structure, complementary to that obtained from generalized
parton distribution functions.
For  the final state relevant to this analysis, \textsl{i.e.} four-jet production in $pp$ collisions,
recent { experimental} results are in the range
{  $\sigma_{eff} \sim 8-35$ mb \cite{4jet,CMS:2021ijt}.}


We propose here a strategy to extract {novel} information on the partonic structure of the proton by considering DPS processes in photon-proton interactions.
{ In such a process, the impact of MPI has been studied in Ref. 
\cite{Butterworth:1996zw} 
via Monte Carlo simulations whereas the DPS contribution has been considered in Ref. \cite{Blok:2014rza}
in the direct photon kinematics, 
in which the DPS processes is initiated  
by a $c \bar c$ pair originating 
from the perturbative splitting of the photon.}
It is well known that in high-energy reactions a quasi-real photon exhibits a rather complex hadronic structure {\cite{Klasen:2002xb}}.
It can interact as a point-like particle with partons in the hadronic 
target, but it can also resolve into a hadronic structure and its 
partonic constituents could participate in the hard scattering.
Additionally, as far as the (low) virtuality $Q^2$ of the photon 
can be measured by tagging 
 the photon-emitter particles, the average transverse 
size of the $q \bar q$ pair fluctuation, $\langle b_\perp^2 \rangle_\gamma $ can be controlled, since it scales as $ 1/Q^2$ \cite{Frankfurt:1981mk,Nikolaev:1990ja}. Therefore, we {address} the
intriguing question
whether a DPS contributions could be observed 
in quasi-real photon-proton interactions, in full analogy with hadronic ones.
In the present letter we will consider $ep$ collisions where the electron is the photon emitter. The generalization 
to other reactions involving nucleon and/or nuclei, only requires the use of the appropriate photon flux factors.
In such a favourable environment therefore 
the DPS mechanism, which is especially sensitive to { parton pair 
correlations in transverse plane} of the colliding particles, could be 
studied
with a projectile of variable and controllable transverse size.
Since a complete formulation of photon dPDFs capturing its longitudinal 
and transverse structure accommodating both its electromagnetic and 
hadronic
components is missing at the moment, in the present paper we elaborate on a much simpler quantity, $\sigma^{\gamma p}_{eff}$. 

{ With those results at hand, 
we  present a first estimate of the DPS cross section for the photo-production of four-jet in HERA kinematics  accompanied by its main background, \textsl{i.e.} the SPS four-jet 
photo-production cross-section, which, to the best of our knowledge, has never being discussed in the literature.
A reliable estimate of the latter gives, in fact, a limit on the DPS contribution
and thus constrains the corresponding models both of the photon and of the proton.} 
We then show  that if the $Q^2$ dependence of $\sigma_{eff}^{\gamma p}$ could be measured, then 
{ a first estimate of the mean transverse distance between partons in 
the proton could be obtained. }
{Such a procedure avoids the intrinsic limitations 
in the extraction of this quantity from $\sigma_{eff}$ which are discussed in Refs. \cite{rapid,jhepc}.}
Moreover we derived lower limits 
on the necessary { integrated} luminosity to observe the predicted $Q^2$
dependence of the
DPS cross sections { in HERA kinematics.}

\section{Effective cross section for $\gamma p$ DPS }

The $\gamma p$ DPS cross section, 
$\sigma_{DPS}^{\gamma p}$, 
can be written in full analogy 
{with the one appearing} in the $pp$ case \cite{1a}.
The {former} does depend on both 
the proton and photon dPDFs, 
{${D_{q_i q_j/p}}(x_i,x_j,k_\perp)$} and
${D_{q\bar q/\gamma}}(x_k,x_l,k_\perp)$, respectively where $ij$ and $kl$
are the flavours of the interacting partons, $k_\perp$ 
is the momentum imbalance, Fourier conjugate variable to the partonic transverse distance, $b_\perp$, and $x$'s are the longitudinal momentum fractions carried by each parton.   
The photon contribution to the cross section can be formally written similarly to that of meson dPDFs \cite{noipion,Kasemets:2016nio}.  
{ In particular, the Light-Front (LF) wave functions of the photon can 
be generally treated as that of a vector meson. With this respect, the 
dPDFs of the $\rho$ meson have been investigated in Ref. 
\cite{Rinaldi:2020ybv}.
{We leave for future analyses the study 
of the rich spin structure of these vector systems
which could give access to new double spin correlations in the proton.} }  

The pair production amplitude, at a given photon virtuality $Q^2$,  can be described within a Light-Front formalism in terms of the LF wave function, $\psi^{ \gamma}$ \cite{noipion}:
\begin{align}
\label{gdpdf}
{D_{q\bar q/\gamma}}(x,\vec k_\perp;Q^2)&=\int d^2 \vec k_{\perp,1} \; \psi_{q \bar q}^{\dagger \gamma}(x, \vec k_{\perp,1};Q^2)
\\
\nonumber
&\times \psi_{q \bar q}^{ \gamma}(x, \vec k_{\perp,1}+ \vec k_\perp;Q^2)~.
 \end{align}
In the above equation we take into account the lowest Fock components.  
This in turn implies that, being a two-particle state, 
the longitudinal momentum of the second parton is given by $1-x$. The integration runs over 
the intrinsic transverse momentum of one parton of the pair, $\vec k_{\perp,1}$, with $\vec k_{\perp,2}=- \vec k_{\perp,1}$.

Given the LF description of the unpolarized  dPDFs and PDFs \cite{noij1,noipion}, one can derive the expression of the effective cross section in terms of effective form factors (effs) \cite{noiPLB}.
The latter, for the photon, reads: 
\begin{align}
    F_2^{\gamma} (\vec k_\perp) = 
  \dfrac{ \displaystyle  \sum_{ q}  \int ~dx~ {D_{q\bar q/\gamma}}(x,\vec k_\perp) }{\displaystyle  \sum_{q} \int ~dx~ {D_{q\bar q/\gamma}}(x,\vec k_\perp=0) }~,
\end{align}
in which the summation and integration,
at variance with proton case, run
over the indexes of only one parton of the pair.
Such a definition of the eff relies
on the approximation, frequently assumed in 
phenomenological analyses in the $pp$ case,
that momentum correlations and parton flavor dependence are neglected. 
{ Moreover it guarantees that 
$F_2^\gamma (\vec{k}_\perp=0) = 1$,
as required by the probabilistic interpretation of 
double parton distributions {in coordinate space} and their corresponding normalization.}
In terms of these quantities, the $\gamma p$
effective cross sections can be written as:
\begin{align}
    \label{sigmaeff1}
    \sigma_{eff}^{\gamma p}(Q^2) =  \left[ 
\int \dfrac{d^2k_\perp}{(2 \pi)^2} F_2^p(k_\perp) F_2^\gamma (k_\perp;Q^2) \right]^{-1}\,.
\end{align}
{  Under the additional assumption that 
double PDFs can be written, at any { perturbative} scale, as
product of ordinary PDFs, the $\gamma p$ DPS cross section for the production of the final state $A+B$ 
is rearranged  in a pocket
formula $\sigma_{DPS}^{A+B} \sim \sigma_{SPS}^A  
\sigma_{SPS}^B/\sigma_{eff}^{\gamma p}$.
Such an approximation allows an estimation of 
$\sigma_{DPS}$ by making use 
of known calculations of single parton scattering (SPS) cross sections $\sigma_{SPS}^{A(B)}$ with $A(B)$ final states.
It is worth to remark that such a procedure, largely used in DPS in $pp$ collisions and also adopted here,  
neglects any type of perturbative and non-perturbative correlations in double PDFs.}


\section{Numerical results for $\sigma_{eff}^{\gamma p}$}
\label{IV}
The evaluation of  
$\sigma_{eff}^{\gamma p}(Q^2)$ requires the knowledge of the proton effective form factor
for which  we use phenomenological 
parametrizations. In particular 
we consider the dipole one of Ref. \cite{Blok2}, which we address as model ``S'':
\begin{align}
    \label{strik}
      F_2^p(\vec{k}_\perp) = \left(1+{k_\perp^2}/m_g^2    \right)^{-4}~\,,
\end{align}
with $m_g^2$=1.1 Ge$\mbox{V}^2$. 
Such a model returns a $\sigma_{eff}^{pp}\sim 30$ mb. 
In addition, in order to explore the dependence of our results on the functional dependence of the proton eff, 
we also considered a Gaussian ansatz \cite{Calucci:1999yz} of the type \begin{align}
    \label{eff:G}
    F_2^p(\vec{k}_\perp)&=e^{-\alpha_i k_\perp^2 },
    \; \; \; \;\; \;i=1,2 \,. 
\end{align}
The parameter $\alpha$ is fixed 
to $\alpha_1=1.53$ GeV$^{-2}$ which returns  $\sigma_{eff}^{pp}=15$ mb (``G$_1$'' model)
and to $\alpha_2=2.56$ GeV$^{-2}$ which returns   $\sigma_{eff}^{pp}=25$ mb (``G$_2$'' model).
All considered proton effs satisfy the normalization condition $F_2^p(\vec{k}_\perp=0)=1$.

The other input appearing in Eq. (\ref{sigmaeff1}) is the photon eff. 
The latter is calculated making use 
of the photon wave functions presented in Refs. 
\cite{,arriola,brodfrank}. 
Among those presented in Ref. \cite{arriola}, we make use of 
the wave functions corresponding to the so called ``spectral quark model''. 
In Ref. \cite{brodfrank}, those
quantities were evaluated to lowest order QED in momentum space.
{ As it is well known \cite{Nikolaev:1990ja}, the modulus squared of the corresponding wave function is logarithmically divergent at large parton transverse momentum $k_{\perp,1}$ in Eq. (\ref{gdpdf}).
 One option to regulate it is by introducing an upper physical cut-off on the 
$k_{\perp,1}$ integration, possibly of the order of the hard scale entering the scattering process.
In the present work we pursued instead the idea of considering a large $k_{cut}$ 
for the reason to be detailed hereafter.
When increasingly higher cut-offs are used
in the evaluation of $F_2^\gamma$, 
its tail at large $k_{\perp}$ 
shows, as expected, a sensitivity to 
$k_{cut}$ and it approaches a constant value of 1 for asymptotically large values of $k_{cut}$, \textsl{i.e.}
the form factor of a structureless photon.
 However, the corresponding variations on 
the $\gamma p$ effective cross-section 
evaluated via Eq. (\ref{sigmaeff1}) are much reduced since the fast falling behaviour of the proton eff at high $k_\perp$ \cite{rapid,jhepc,Blok2}  
effectively regulates the tail of the photon eff in the convolution integral in Eq. (3) and grants its convergence.
Althought a residual dependence 
of $\sigma_{eff}^{\gamma p}$ on $k_{cut}$ 
is still present, we have numerically  verified that the effective cross-section varies no more than $\sim 1$ mb when $k_{cut}$ is raised from $k_{cut}=50$ GeV to $k_{cut}=10^3$ GeV,
with the latter being used as  default value in the following.
{See further details on this topic in the appendix \ref{a2}.}
Given these observations, the advantage of using a large cutoff is twofold: firstly one avoids to introduce a prescription for setting a physical cut-off and the resulting arbitrariness; secondly one obtains a lower limit on the $\gamma p$ effective-cross section, which implies that our estimate of the DPS cross section given by QED contribution should be considered as its upper limit. 
This procedure is analogous
to the one adopted in Ref. \cite{Nikolaev:1990ja} in coordinate space, 
where the square modulus of the photon wave function, divergent 
in the small $b_\perp$ limit, is de facto regularized by the so-called dipole 
cross section which vanishes as $b_\perp$ goes to zero .
In the case of hadronic model of the photon 
presented in Ref. \cite{arriola} such
an issue is not present, since the corresponding wave function is properly normalized from the beginning.

We present our numerical estimates 
for $\sigma_{eff}^{\gamma p}(Q^2)$
in Fig. (\ref{Newfig}).
Since in the present paper we will 
consider photoproduction in $ep$ collisions, 
the lower limit on $Q^2$ is set of the order
of $m_e^2$, the mass of the electron, 
appearing in the Weizsäcker-Williams approximation
for the spectrum of the exchanged photon.
One may notice that the hadronic models 
of Ref. \cite{arriola} returns a 
systematically higher $\sigma_{eff}^{\gamma p}$
with respect to the pure electromagnetic one 
\cite{brodfrank}. The spread between the curves 
pertinent to the same photon model indicates a rather large sensitivity to the proton effective form factor. 
Both models display a peculiar pattern 
of the $Q^2$ dependence: both start from a plateau at low  $Q^2$ and decrease at larger $Q^2$, 
with the onset of the decrease occurring at rather different values of $Q^2$. The shape of the distribution is replicated irrespectively  
of the adopted proton eff. 
We observe that, in the limit of high photon 
virtuality, the value of $\sigma_{eff}^{\gamma p}$ can be predicted in complete analogy with the gluon splitting case elaborated in Ref. \cite{Gaunt:2012dd}. 
In fact, the $q\bar q$ pair, originated by the electromagnetic splitting of a highly virtual photon, is characterized by quite small transverse distance, $\langle b_\perp^2 \rangle_\gamma \propto 1/Q^2$, and asymptotically one has  $(\sigma_{eff,asy})^{-1} = \int d^2k_\perp/(2 \pi)^2~F_2^p(k_\perp)=\tilde{F}_2^p(b_\perp=0) 
$ \cite{Gaunt:2012dd}, { where $\tilde F_2^p(b_\perp)$ is the Fourier 
transform of the eff and  $b_\perp$ is the conjugate variable to 
$k_\perp$}. Adopting the G$_1$ model for the proton eff, the calculation 
with the photon wave function of Ref. \cite{brodfrank} returns a value $ 
\sigma_{eff}^{\gamma p}(Q^2=100~\mbox{GeV}^2) \sim 7.52 $ mb while the 
predictions from Ref. \cite{Gaunt:2012dd} would give  $\sigma_{eff,asy} 
\sim 7.5$ mb, showing remarkable agreement.
{ We point out that the analysis of 
Ref. \cite{Blok:2014rza} is performed in the 
approximation described above, where the perturbative splitting of the photon into a $c\bar{c}$ pair 
probes the proton eff (model S) at zero transverse distance, which makes that analysis complementary 
to the one discussed here, 
where the quasi-real photon develops a partonic structure at larger transverse distances.} 
%
%

\begin{figure}[t]
\hskip 0cm \includegraphics[scale=0.6]{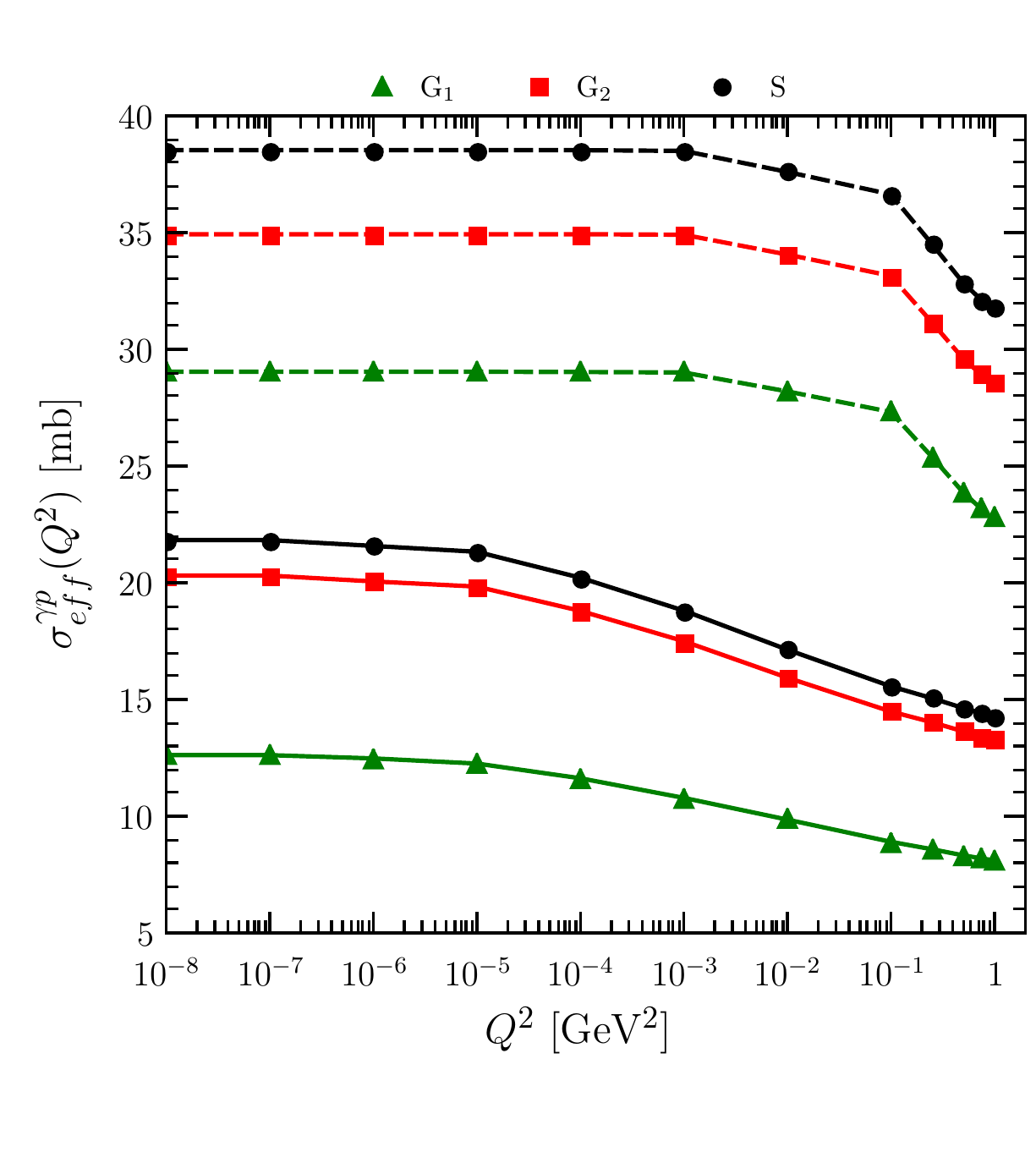}
\vskip -1cm
\caption{\footnotesize $\sigma_{eff}^{\gamma p}$   evaluated in Eq. (\ref{sigmaeff1}) 
 with the w.f. of Ref. \cite{arriola} (dashed lines) and  Ref. 
\cite{brodfrank} (full lines) as a function 
of $Q^2$. Different symbols denote the 
proton effs described in the text.}
\label{Newfig}
\end{figure}

\section{The four-jet photo-production cross-section}

The four-jet final states have been measured in photoproduction at HERA by the ZEUS collaboration \cite{Chekanov:2007ab}. In that analysis, they considered jets with transverse energy 
$E_T^{jet} > 6$ GeV and laboratory pseudorapidity
$|\eta_{jet} | < 2.4$, in the kinematic region  $Q^2 <
1 \; \mbox{GeV}^2$ and  
{the energy fraction transferred from the lepton to the photon, $y=E_\gamma/E_l$, in the range $0.2 \leq y \leq 0.85$}. 
The comparison with leading-logarithmic
parton-shower Monte Carlo models 
\cite{Marchesini:1991ch,Butterworth:1996zw,Sjostrand:2000wi}
showed that the inclusion in the simulation of multi-parton interactions 
significantly improve the description of the data.
This stimulated us to estimate the $\sigma_{DPS}$ { contribution to that 
final state by adopting} kinematical cuts of Ref. 
\cite{Chekanov:2007ab}.
The expression for the differential $\sigma_{DPS}$ 
{initiated by a quasi-real photon
is then generalized
according to the results presented 
in Ref. \cite{Gaunt:2012dd}}:
\begin{align}
\label{sigmaDPS}
d\sigma_{DPS}^{4j} = & \frac{1}{2}
\sum_{ab,cd} \int dy \; dQ^2 \;
\frac{f_{\gamma/e}(y,Q^2)}{\sigma_{eff}^{\gamma 
p}(Q^2)} \times  \\
&\times \int dx_{p_a} dx_{\gamma_b} 
f_{a/p}(x_{p_a}) f_{b/\gamma}(x_{\gamma_b})
d\hat{\sigma}^{2j}_{ab}(x_{p_a},x_{\gamma_b})
\nonumber \\
&\times \int dx_{p_c} dx_{\gamma_d} 
f_{c/p}(x_{p_c}) f_{d/\gamma}(x_{\gamma_d})
d\hat{\sigma}^{2j}_{cd}(x_{p_c},x_{\gamma_d})\,.
\nonumber
\end{align}
In the above equation the one half factor takes into account two identical dijet 
systems in the final state. 
The sum runs over partons active in the first scattering  
($a,b$) or in the second one ($c,d$), where 
$d\hat{\sigma}^{2j}$ represent the differential partonic cross sections.
Since  $\sigma_{eff}^{\gamma p}$ depends on $Q^2$, the photon flux, $f_{\gamma/e}$, in its
$Q^2$-unintegrated version  \cite{Frixione:1993yw}, has been used. 
The distributions $f_{i/p}(x_{p_i})$ and $f_{j/\gamma}(x_{\gamma_j})$ 
represent the proton and of the photon PDFs
for which we use the leading order sets 
of Ref. \cite{Pumplin:2002vw} and \cite{Gluck:1991jc}, respectively.
Dijet cross sections have been evaluated to leading order accuracy in the strong coupling
{ with \texttt{ALPGEN} \cite{Mangano:2002ea}, properly adapted to cope 
with photo-production processes},
with final state partons identified as jets. Factorization and renormalization scales have been both set to average transverse  momentum of the jets.
Such a cross section receives two contributions that can be classified by the fractional momentum of partons in the photon, $x_\gamma$, reconstructed by jet kinematics:
the one from the resolved photon process, in which 
the photon behaves like an hadron with its own parton distributions, 
and the direct one, in which the photon interacts as a point-like 
particle
with partons in the proton target. The former populates the whole 
$x_\gamma$ range while the latter, at LO, is concentrated at $x_\gamma=1$.
The two mechanisms mix under { higher-order} corrections \cite{Frixione:1997ks}
and therefore
kinematic cuts are used by experimental collaborations 
{ \cite{ZEUS:2007njl,H1:2006rre}}
to select the 
resolved-enriched contribution { ($x_\gamma<0.75$) and a direct-enriched 
one ($x_\gamma>0.75$). 
In the present analysis we are interested in the resolved component and therefore the cut $x_\gamma<0.75$ is enforced on the evaluation of dijet cross sections.
Out of this predictions, the DPS cross section
is built via the pocket formula in Eq. (\ref{sigmaDPS}) by enforcing, for consistency, the same cut on the parton pair fractional momenta, $x_{\gamma,1}+x_{\gamma,2}<0.75$.
The main background to the DPS signal is represented by the SPS four-jet photoproduction process. 
The latter is again calculated with \texttt{ALPGEN}
interfaced with the photon flux factor and photon PDFs, enforcing $x_\gamma<0.75$ 
and with same settings discussed for the dijet cross sections. 
The experimental four-jet photoproduction cross section, {($\sigma_{exp}$)}, that can be inferred from distributions presented in Ref. \cite{Chekanov:2007ab} for $x_\gamma<0.75$ is 135 pb.}

\begin{table}[h]
\begin{center}
\begin{tabular}{ccccccc}
\hline
\hline
     \rule{0mm}{0.4cm} 
 & & $  Q^2 \leq 10^{-2}$  & $10^{-2}\leq Q^2 \leq 1$ & $Q^2 \leq 1$  &$  
\frac{\displaystyle \sigma  }{\displaystyle \sigma_{exp}}$& $R$ \\
  \rule{0mm}{0.4cm}
  & & [GeV$^2$]  & [GeV$^2$] & [GeV$^2$]  & [\%] & \\  \hline
     \multirow{2}{*}{   }& & \multicolumn{4}{c}{$\sigma_{DPS}$ [pb]} 
     \rule{0mm}{0.4cm}
      \\ \hline
\multirow{3}{*}{~w.f. }& \mc{1}{|l|}{G$_1$} & 35.1& 
18.6 & 53.7 &40 & 1.89  \\
& \mc{1}{|l|}{G$_2$}  & 29.1              
   &  15.2 & 44.3 & 33 & 1.91 \\
\cite{arriola} & \mc{1}{|l|}{S}  &   26.4                   & 13.7
& 
40.1 & 30 & 1.93  \\  \hline
\multirow{3}{*}{~w.f. }& \mc{1}{|l|}{G$_1$} & 87.8& 
54.3 & 142.1&   101 & 1.62  \\
& \mc{1}{|l|}{G$_2$}  &  54.3                
   &  33.4 & 87.7 &65  & 1.63 \\
\cite{brodfrank}& \mc{1}{|l|}{S}  &   50.5& 31.1 & 81.6 & 60 & 1.62  \\
\hline
\mc{2}{l}{}& \multicolumn{4}{c}{$\sigma_{SPS}$ [pb]} 
     \rule{0mm}{0.4cm}
 \\ \hline
 \mc{2}{l|}{~~~LO SPS} & 77.5& 
36.6 & 114.1 & 86 & 2.12  \\
\hline
\hline 
\end{tabular}
\caption{\footnotesize Predictions for the LO DPS and SPS cross sections for four-jet photo-production in three ranges of $Q^2$. In the last column, the ratio between the calculated cross-sections  to the total one is displayed.
In the DPS case, each row corresponds to prediction obtained with a given $pp$ eff ($G_1$,$G_2$,$S$), 
and the photon wave function of Refs. \cite{arriola}
 (three upper rows) and Ref.\cite{brodfrank} (three bottom rows). In the last column the ratio Eq. (\ref{ratio}) is shown. }
\label{tab2}
\end{center}
\end{table}

We report in Tab. (\ref{tab2}) the results for the $\sigma_{DPS}$ and $\sigma_{SPS}$ obtained for three ranges of photon virtualities in HERA kinematics.
Predictions are displayed on different row 
depending on the adopted proton eff
and photon wave functions.
As far as the comparison with the experimental 
cross section for $Q^2<1$ Ge$\mbox{V}^2$ is concerned,
the DPS cross section gives a sizeable contribution
for all configurations { whereas the {LO} SPS almost saturates
the 
experimental cross section. }
{ With this respect, this preliminary investigation already indicates that some
configurations, e.g. the LO QED description of the photon combined with the
$G_1$ proton eff, are not favorable combinations since the corresponding DPS
cross section alone exceeds the experimental one. }

{
These results, however, should be interpreted with special care. 
Higher order corrections to the 
dijet photoproduction cross section
induce an increase of theoretical predictions by a factor of 1.3 going from LO to NLO \cite{Klasen:1996it,Klasen:1997br}
and by a factor 1.05 going from  NLO to NNLO \cite{Klasen:2013cba}.
This in turn implies that by using LO 
estimates for the dijet cross section
in the pocket formula, the latter 
gives a lower limit on the DPS cross section, as far as { higher-order}
corrections are considered.
On the other hand, a good theoretical control 
of { higher-order} corrections to the SPS background
is mandatory for a proper extraction of
DPS signal. For example, the large spread in $\sigma_{eff}$ values 
reported in the experimental analysis of Ref. \cite{CMS:2021ijt}
reflects the level of uncertainty in the theoretical estimation of the $\sigma_{SPS}^{4j}$. 
With this respect, NLO results for four-jet production in $pp$ collisions at 8
TeV have been first calculated in Ref. { \cite{Badger:2012pf,Bern:2011ep}}
showing that NLO predictions are nearly half 
of the LO estimates. If such a trend should be confirmed also in four-jet 
photo-production in HERA kinematics, our LO result would represent therefore an upper limit on the SPS background. 
The scenario concerning 
the uncertainties connected to higher order corrections appears as follows. Our LO estimates sets a lower limit
on the DPS cross section, whereas we have presented arguments showing that the LO SPS 
could represent its upper bound. Both these findings converge into a conservative scenario for the DPS contribution to four jet cross section.
For those reasons we do not provide here 
a full propagation  of theoretical uncertainties to our final results. 
Nevertheless, it is  
worth to mention that the largest theoretical uncertainty, and by far dominant over all others, comes from a) models of the proton structure and the spread in the corresponding  $\sigma_{eff}^{pp}$ values and b) the use of a LO QED treatment of the photon in addition to quark models for its hadronic component, without their consistent combination into a photon double PDFs, which in turns
generates a wide spread on  $\sigma_{eff}^{\gamma p}$ predictions. }


Despite all these warnings, 
all models predict large DPS fractions 
suggesting that jets photoproduction 
in $ep$ collisions could represent an interesting channel to search for the DPS contribution.


\begin{figure}[t]
\includegraphics[scale=0.6]{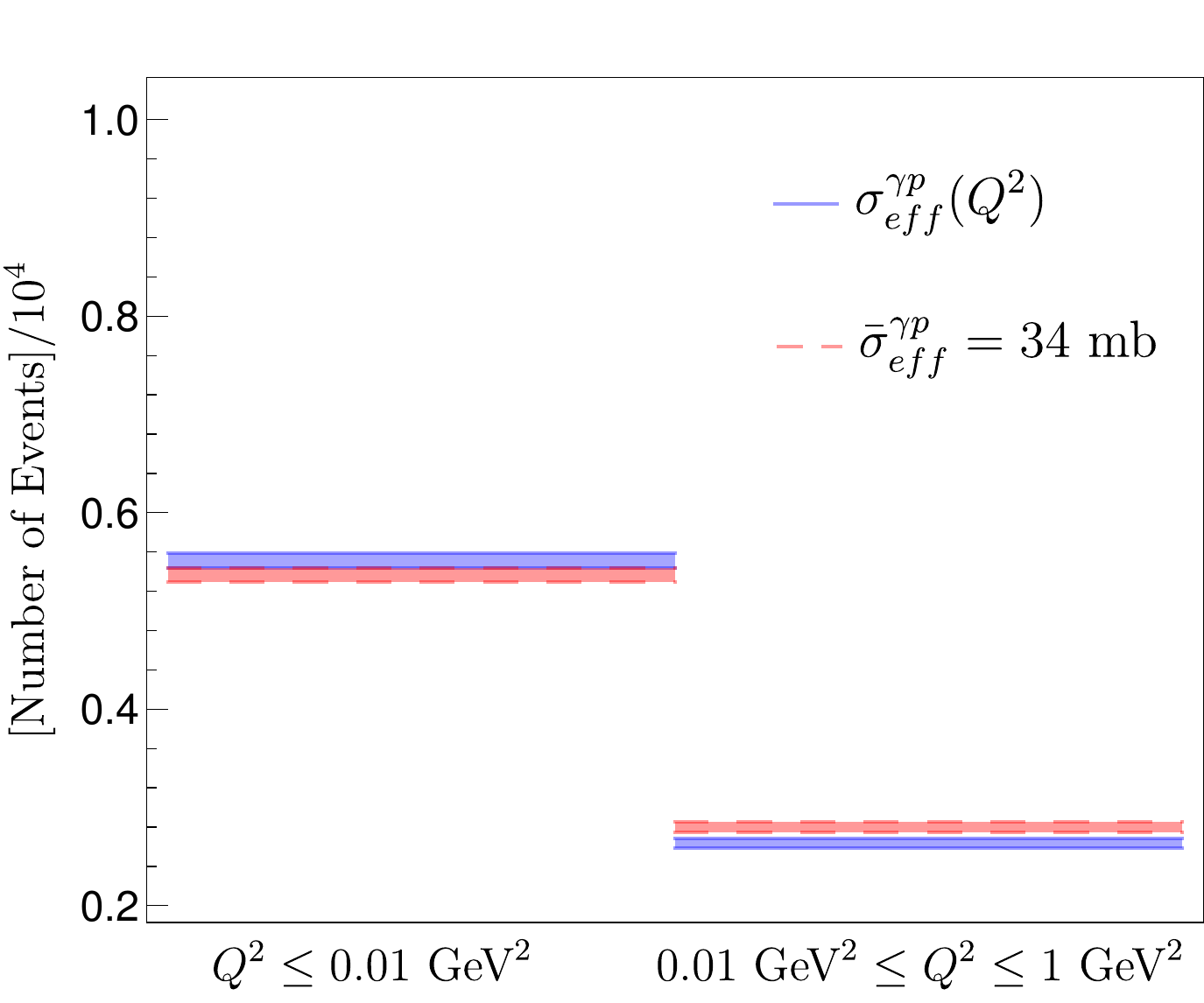}
\caption{\footnotesize The estimated number of events as a 
function  of $Q^2$ for 200 $pb^{-1}$ of integrated luminosity for the photon model of Ref. \cite{arriola} and proton eff G2.
 Full lines stand the  evaluations of $\sigma_{DPS}$ by means of  $\sigma_{eff}^{\gamma p}(Q^2)$ and  
the 
dotted ones represent the calculations of $\sigma_{DPS}$ by using the  
$Q^2$-independent $\bar \sigma_{eff}^{\gamma p}$. }
\label{Nev_vs_Q2}
\end{figure}

\section{Extraction of the $Q^2$-dependence of $\sigma_{eff}^{\gamma p}$}

{ 
Given the rather large uncertainties 
on the absolute DPS cross sections, it seems to us premature to consider 
more differential observable at this point.
We make only one exception by discussing 
the $Q^2$-dependence of the DPS cross section.
The latter is of primary importance to us since it is linked to the concept of a photon of variable transverse size and, in turn, to a $Q^2$-dependent $\sigma_{eff}^{\gamma p}$.  
We investigate whether such a dependence, which adds on top of the one naturally induced by the photon flux, is eventually observable.  
We perform such an analysis within the HERA settings presented in the previous Section.
We set the notation by sketching Eq. (\ref{sigmaDPS}) as $ d\sigma_{DPS}(bin) \sim \int_{bin}dQ^2 g(Q^2)/\sigma_{eff}^{\gamma p}(Q^2)$, where $bin$ stands for a given interval of  integration over $Q^2$ and the function $g$ encodes the flux factor, the PDFs and elementary cross sections. 
We present in the first two columns of 
Tab. (\ref{tab2}) the DPS cross section
integrated over ranges of photon virtualities.
Then we define the ratio $R$:}
\begin{align}
  R= \dfrac{ d\sigma_{DPS}(bin1)}{d\sigma_{DPS}(bin2)}~.  
  \label{ratio}
\end{align}
{In the case $\sigma_{eff}^{\gamma p}$ were a constant,  the latter 
quantity would be: $R \sim \int_{bin1}dQ^2 g(Q^2)/ \int_{bin2}dQ^2 
g(Q^2) \sim 2.1$. Therefore, if the ratio of the DPS cross sections, 
evaluated in the two bins,  
results to be different from that number,
this fact would directly  point to $Q^2$ effects on $\sigma_{eff}^{\gamma p}$ or possible correlations breaking the pocket formula. 
As one can see in the last column of Tab. (\ref{tab2}), LO QED and the model encoding non perturbative QCD effects predict deviations of $R$ from the reference value of $2.1$. 
We close this section remarking that this ratio is particular effective since it does not depend on the chosen proton and photon effs, it is sensitive to the dependence of  $\sigma_{eff}^{\gamma p}$ on $Q^2$ and finally, if applied to the SPS four-jet
cross section, its value gives a theoretical benchmark without requiring the exact knowledge of the absolute four-jet cross section. }

{ Furthermore, we also developed a procedure to establish the minimum {
integrated}
luminosity  to experimentally access $Q^2$ effects in 
$\sigma_{eff}^{\gamma p}(Q^2)$. 
We have converted the cross sections in Tab. (\ref{tab2}) in expected number of
events with a given { integrated} luminosity. Statistical errors and the
corresponding bands are calculated assuming  a Poissonian distribution.
The results are presented in Fig. (\ref{Nev_vs_Q2}) where the blue curves indicate results for the
$\sigma_{eff}^{\gamma p} (Q^2)$ 
while red ones indicate the number of events obtained with a constant, $Q^2$ independent, $\bar \sigma_{eff}^{\gamma p}$ which reproduces the total cross section
for $Q^2<1$ Ge$\mbox{V}^2$ obtained with $\sigma_{eff}^{\gamma p}(Q^2)$.
Among the various model results listed 
in Tab. (\ref{tab2}), in Fig. (\ref{Nev_vs_Q2}) 
it is shown, on purpose, the one with the lower cross section and with a smooth $Q^2$ dependence.
We find that the minimal { integrated} luminosity which makes
distinguishable the two models giving non overlapping error bands and therefore  
exposes the $Q^2$-dependence of $\sigma_{eff}^{\gamma p}$ is 
$\mathcal{L}=200$ pb$^{-1}$.}

{ This { integrated} luminosity estimate assumes that the
SPS background could be subtracted with high efficiency 
and therefore it should be considered as a 
lower limit to start observing $Q^2$ effects in $\sigma_{eff}$.}

\section{The geometry of 
 $\sigma_{eff}^{\gamma p}(Q^2)$ }
\label{III}

As discussed in Refs. \cite{rapid,jhepc},  the rather limited knowledge of the proton 
eff, entering the definition of $\sigma_{eff}^{pp}$, prevents a
precise extraction of  
${\langle b^2_\perp \rangle_p}$.
{ In fact, in $pp$ or $pA$ collisions, the integrand defining the 
relative $\sigma_{eff}$, see e.g. Eq. (\ref{sigmaeff1}),
is completely unknown. Therefore, without assuming some model constraints, one cannot directly 
relate  ${\langle b^2_\perp \rangle_p}$ to experimental data. }
The possibility to have in $\gamma p$ interactions a projectile of variable size, depending on $Q^2$, can provide a unique chance
to extract ${\langle b^2_\perp \rangle_p}$.
{Without specifying any peculiar photon w.f.,}
one may define the Fourier Transform of the eff, $\tilde F_2(b_\perp)$, which is interpreted as the probability distribution of finding two partons at a given transverse distance $b_\perp$  
\cite{Calucci:1999yz,rapid,jhepc}. 
{ Once this quantity has been} evaluated within some model of the photon 
structure, it can be power expanded as:
\begin{align}
    \label{exp1}
    \tilde F^\gamma_2(b_\perp;Q^2)= \sum_n~C_n(\bar b_\perp;Q^2)(b_\perp-\bar b_\perp)^n~,
\end{align}
and freedom is left in the choice in the expansion point $\bar b_\perp$.
Eq. (\ref{sigmaeff1}) can now be  rewritten as:

\begin{align}
    \label{rel1}
   \Big[ \sigma_{eff}^{\gamma p}(Q^2) \Big]^{-1} &=
 \int d^2 b_\perp~ \tilde F^p_2(b_\perp) \tilde F^{\gamma}_2(b_\perp;Q^2)
   \\
   \nonumber
   &= \sum_n C_n(\bar b_\perp;Q^2) \langle (b_\perp-\bar b_\perp)^n \rangle_p.
\end{align}
A realistic description of $C_n(\bar b_\perp;Q^2)$, together with  data on the $Q^2$ 
dependence of $ \sigma_{eff}^{\gamma p}(Q^2) $, will allow to
 access  the transverse distance of partons in the proton. 
 In fact, {for a given specific dependence of $C_n$ on $Q^2$, one can identify an}
 operator, $\mathcal{O}_{Q^2}^m$, such that
\begin{align}
\label{ope}
   \mathcal{O}_{Q^2}^m  \Big[ \sigma_{eff}^{\gamma p}(Q^2) \Big]^{-1} = 
 \mathcal{O}_{Q^2}^m C_m(\bar b_\perp,Q^2)  \langle (b_\perp-\bar b_\perp)^m \rangle_p\,,
\end{align}
{and then } one {can select}  and extract
 $ \langle (b_\perp-\tilde b_\perp)^m 
\rangle_p$, i.e. the
relevant information on the proton structure. {For the moment being we do not specify any functional expression of $ \mathcal{O}_{Q^2}^m $.}
{ This quantity, related to the explicit expressions of $C_m(\bar 
b_\perp,Q^2)$, could be, e.g., a proper differential operator on $Q^2$. 
}
We have successfully tested the procedure both analytically and numerically with different proton and photon eff models. The identification of the correct operator is however not 
unique and freedom is left in the choice of the expansion point.
Such a flexible feature can be useful 
for possible experimental applications.
The only practical limitation is represented by the accuracy with which 
the dependence of $\sigma_{eff}^{\gamma p}$ on $Q^2$ could be eventually measured.
{ Therefore, this procedure should be properly optimised along the 
experimental extraction conditions. In closing this section, we stress 
again that the procedure can be used only by considering realistic 
description of the photon splitting mechanism by taking into account 
higher order QED effects. 
{Examples of application of this procedure are discussed
in the appendix \ref{a1}.}

This relation is one of the main goals of the present analysis and constitutes  motivation to suggest this type of  measurements at facilities where the photon virtuality can be experimentally measured such as the future  Electron Ion Collider \cite{AbdulKhalek:2021gbh}. }

\section{Summary}
 In the present analysis we have derived 
effective cross sections 
for photon induced processes
which are essential ingredients in the predictions
of DPS cross sections in quasi-real photon proton interactions. The latter have been obtained 
with the help of electromagnetic and hadronic 
model of the photon formulated in terms 
of light cone wave functions.
For the four-jet final state in HERA kinematics we found a sizeable DPS contribution. 
{ This conclusion persists after 
considering estimates of higher order corrections, both to the DPS and SPS processes, taken from the literature.}
In the case the photon virtuality $Q^2$ could be measured, we have investigated the dependence
of $\sigma_{eff}^{\gamma p}$ 
on such a parameter, which is directly related
to the size of the dipole originated by the photon fluctuation. We set lower limits 
on the integrated luminosity needed to observe such an effect and {we present, for such a case, a novel procedure which would allow to extract new information on the proton structure.}

This work was supported: {\it i)} in part by  the STRONG-2020 project of the
European { Union} Horizon 2020 research and
innovation programme under grant agreement No 824093;
{\it ii)} by the European Research Council  under the European Union’s Horizon 2020 research and innovation program 
(Grant agreement No. 804480);  {\it iii)} by the project “Photon initiated double parton
scattering: illuminating the proton parton structure” on the FRB of the University of Perugia.

\newpage
\onecolumngrid
\appendix

\section{ Cut-off dependence of the results} \label{a2}
In this section we discuss the dependence of the outcomes of the present
analysis on the values of $k_{cut}$.
Let us remind that
the modulus squared of the photon QED wave function is logarithmically
divergent at large parton transverse momentum $k_{\perp,1}$ in Eq. (1).
Therefore, a regulating procedure is required. In turn, it is important
to discuss the impact of such a strategy on the results of the calculations.
In fact, several possibilities are available, e.g., we can integrate
up to a physical cut-off, possibly of
the order of the  hard scale entering the scattering process { or related to
other prescriptions}.
In the present work we pursued instead the idea of considering a large
$k_{cut}$
for the reason to be detailed hereafter.
When $F_2^\gamma$ is evaluated with increasingly higher cut-offs, as shown in
Fig. \ref{eff},
its tail at large $k_{\perp}$
shows, as expected, a sensitivity to
$k_{cut}$ and it approaches a constant value of 1 for asymptotically large
values of $k_{cut}$, \textsl{i.e.} the form factor of a structureless photon.

\begin{figure*}[h]
\begin{center}
\includegraphics[scale=0.5]{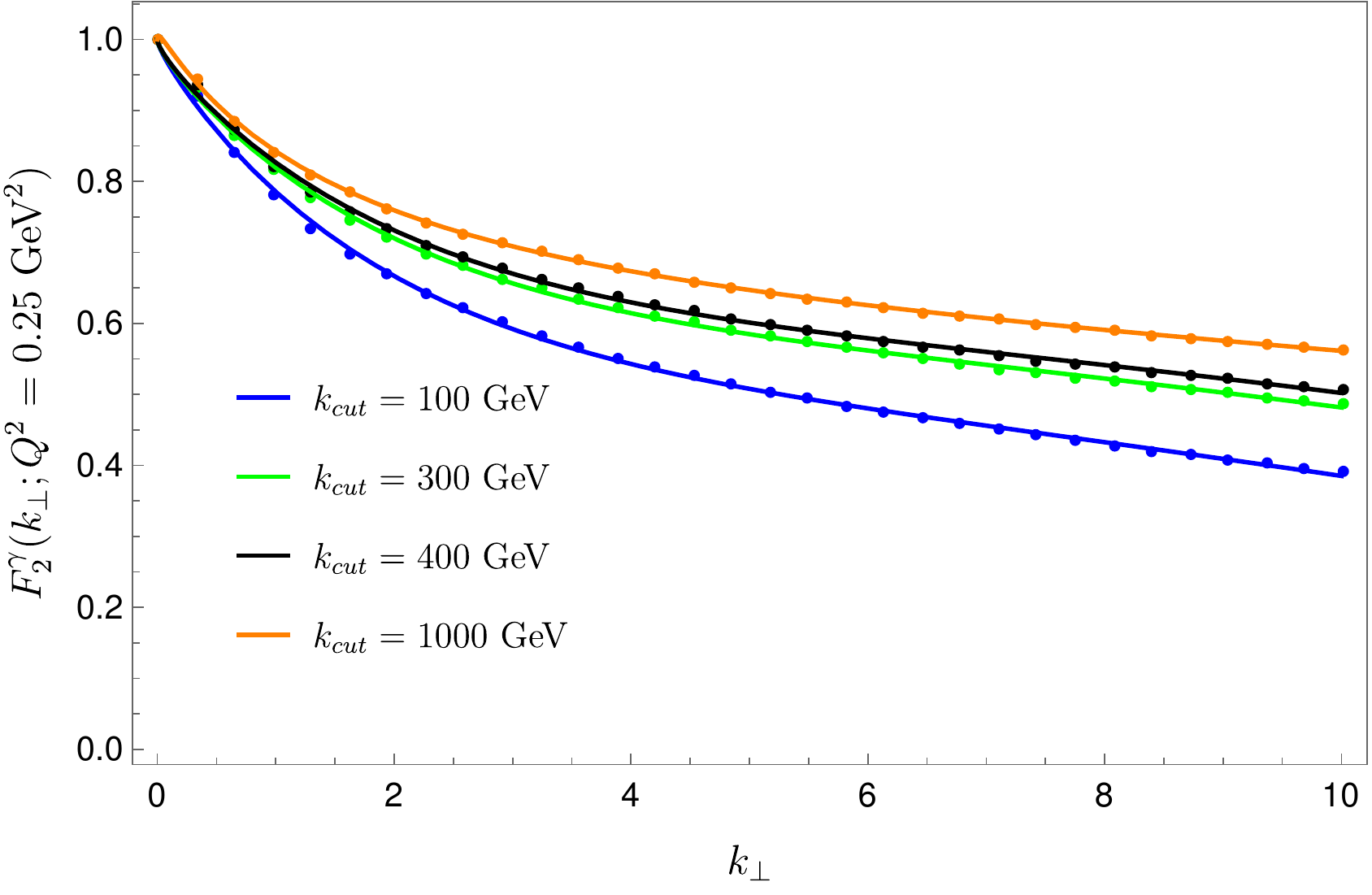}
\caption{The photon eff as a function of $k_\perp$ and evaluated at $Q^2=0.25$
GeV$^2$ for different cut-off.
}
\label{eff}
\end{center}
\end{figure*}

The corresponding variations on
the $\gamma p$ effective cross-section
are instead much reduced and presented in Table \Ref{si_QED} for three proton
models as a function of $k_{cut}$. Its lowest value is in the range of jet
transverse momenta discussed in the phenomenological session, while the maximum
is a numerical proxy for infinity.

\begin{table}[h]
\begin{center}
\begin{tabular}{|c c|c|c|c|c|c|c|c|c|c|}
\hline
     \multirow{2}{*}{   }& & \multicolumn{8}{c|}{$k_{cut}$ [GeV]} \\
\hhline{~~--------}
 &  & 5&10&50&100  &1000&$\Delta_{5-10^3}$ [mb]&$\Delta_{10-10^3}$
[mb]&$\Delta_{50-10^3}$[mb] \\
\hline
\multirow{3}{*}{$\sigma_{eff}^{\gamma p}(k_{cut})$}&
\mc{1}{|l|}{G$_1$} &11.13&10.32&9.16 &8.95   &8.49&2.64&1.87&0.67 \\
\hhline{~---------}          & \mc{1}{|l|}{G$_2$}              &
16.93&16.46&14.63&14.44 &13.85&3.08&2.61&0.78 \\
\hhline{~---------}          & \mc{1}{|l|}{S}       &18.57&17.87&15.81&15.8&
14.89&3.65&2.98&0.92 \\
\hline
\end{tabular}
\caption{The photon-proton effective cross-section, in [mb],
 for $Q^2=0.26$ GeV$^2$  and different
$k_{cut}$ and ansatz for the proton eff. Here we define
$\Delta_{k1-k2}=\sigma_{eff}^{\gamma p}(Q^2,k_1)-\sigma_{eff}^{\gamma
p}(Q^2,k_2)$. }
\label{si_QED}
\end{center}
\end{table}

In all considered cases we stress that
the fast falling behaviour of the proton eff at high $k_\perp$ effectively
regulates the tail of the photon eff in the convolution integral in Eq. (3),
granting its convergence.
As expected, a residual dependence
of $\sigma_{eff}^{\gamma p}$ on $k_{cut}$
is visible and quantified as a difference in the last columns of the Table.
However, we notice that for increasingly higher $k_{cut}$ the values of
$\sigma_{eff}^{\gamma p}$
stabilize: it varies no more than $\sim 1$ mb when $k_{cut}$ is raised from
$k_{cut}=50$ GeV to $k_{cut}=10^3$ GeV. The latter value is therefore used as a
default in the following numerical evaluations.

Given these observations, the advantage of using a large cutoff is twofold: 1)
one avoids to give a particular prescription for the setting of a physical
cut-off 2) one obtains the lower limit on the $\gamma p$ effective-cross
section, which implies that our estimate of the DPS cross section given by QED
contribution should be considered as its upper limit. { We also notice that,
even considering physical cut-offs, we obtain
reduced but still sizeable DPS contributions,
leaving unchanged the main message of the analysis.}

\section{ Examples of the  application of the procedure
Eq. (10)} \label{a1}

In the first part of this additional document, we
provide
examples of the application of the
procedure developed in Sect. VI of the manuscript.
For the { sake} of clarity, ``eq.'' refers to an equation present in this
document while ``Eq.'' refers to that shown in the main manuscript.

We start from Eq. (9)
\textcolor{black}{and without losing} generality,
we consider \textcolor{black}{the expansion point}
$\bar b_\perp=0$ and set
$b_\perp \equiv b$.

\begin{align}
    \Big[\sigma_{eff}^{\gamma p} (Q^2) \Big]^{-1} = \sum_n C^\gamma_n(Q^2) \int
d^2 b~ b^n F_2^p(b) =
\sum_n C_n(Q^2) \langle b^n \rangle_p;
\end{align}
where here $\langle b^n \rangle_p$ is the
\textcolor{black}{$n^{th}$-moment}
of transverse distance between
two partons inside the proton
{and it does not depend
on $Q^2$}.

If a realistic calculation of  $C_n(Q^2)$
is
available,   an operator, depending on $Q^2$, could be identified such that,
 for example:
\begin{align}
    \mathcal{\hat O}^n_{Q^2} ~\Big[\sigma_{eff}^{\gamma p}(Q^2)\Big]^{-1}
\propto C_n(Q^2)
\mathcal{\hat O}^n_{Q^2}~ \langle b^n \rangle_p~.
\end{align}
In other words, it could be possible to select the $n$ power contribution
of the expansion and extract the relevant information on the mean partonic
transverse distance.
In the following we provide two basic examples of the application of this
procedure.

\subsection{Gaussian-Gaussian scenario}

Let us assume that the photon effective from factor
has a Gaussian form of the type:
\begin{align}
    \tilde F_2 ^\gamma(b;Q^2)=\dfrac{Q^2}{\alpha^2 \pi} e^{-b^2 Q^2/\alpha^2} ~,
\end{align}
which is properly normalized:
\begin{align}
    \int d^2b~  \tilde F_2^\gamma(b;Q^2)=1\,,
\end{align}
and whose width is controlled by the
adjustable parameter $\alpha$.
For such a distribution the main transverse distance between the partons
produced by the splitting mechanism is:
\begin{align}
   \langle b^2 \rangle_\gamma= \int d^2b ~b^2 \tilde F_{2}^\gamma(b;Q^2)=
\dfrac{\alpha^2}{Q^2}~.
\end{align}
{Notably,
for a highly virtual photon, i.e. at large $Q^2$, the mean distance between the
two produced partons goes to zero, as expected.  }

Within this choice, we can expand $\tilde F_2^ \gamma(b;Q^2)$ as
\begin{align}
    \tilde F_{2}^ \gamma(b;Q^2) \sim  \dfrac{Q^2}{\pi \alpha^2}-\dfrac{Q^4}{\pi
\alpha^4} b^2+\mathcal{O}(b^4),
\end{align}
where $C_0(Q^2)=
\dfrac{Q^2}{\pi \alpha^2}$ and $C_2(Q^2) =- \dfrac{Q^4}{\pi \alpha^4}$. In this
scenario,
one can chose the relevant operator which isolates
$\langle b^2 \rangle_p$: $\mathcal{\hat
O}= d/(Q^3 dQ)|_{Q^2=0}$. With this choice, we can prove that:
\begin{align}
    \frac{d}{Q^3 d Q} \Big( [\sigma_{eff}^{\gamma p}(Q^2)]^{-1}-C_0(Q^2) \Big)
\Big|_{Q^2=0} =   \frac{d}{Q^3 d Q} \Big( C_2(Q^2) \Big)  \Big|_{Q^2=0}
\langle b^2 \rangle_p~.
\end{align}
Another possibility could be  $\mathcal{\hat O}= d^4/(d^4Q)|_{Q^2=0}$.
In order to show how the procedure works, we further assume that the proton eff
has the form
\begin{align}
\tilde F_2^p(b)= e^{ -b^2 \beta^2  }\dfrac{\beta^2}{\pi}  \,,
\end{align}
which obeys the normalization condition:
\begin{align}
          \int d^2b~  \tilde F_{2}(b)=1\,,
    \end{align}
and for which the partonic mean transverse distance reads
    \begin{align}
          \int d^2b~  b^2 \tilde F_{2}(b)= \dfrac{1}{\beta^2}~.
    \end{align}
From these models of the photon and the proton effs, we may calculate the
$\gamma p$ effective cross section as:
\begin{align}
   [\sigma_{eff}^{\gamma p}(Q^2)]^{-1} =  \dfrac{{\beta}^2 Q^2}{\pi
\left(\alpha^2
\beta^2+Q^2\right)}\,,
\end{align}
and one gets:
\begin{align}
     \frac{d}{Q^3 d Q} \Big( [\sigma_{eff}^{\gamma p}(Q^2)]^{-1}-C_0(Q^2) \Big)
\Big|_{Q^2=0} &= -\frac{4}{\pi  {\alpha}^4 {\beta}^2} \,,
     \\
      \frac{d}{Q^3 d Q} \Big( C_2(Q^2) \Big)  \Big|_{Q^2=0}
&=-\frac{4}{\pi  {\alpha}^4 } \,,
\end{align}
which can be combined to give
\begin{align}
    \langle b^2 \rangle_p =  \dfrac{\frac{d}{Q^3 d Q} \Big(
[\sigma_{eff}^{\gamma p}(Q^2)]^{-1}-C_0(Q^2) \Big) \Big|_{Q^2=0} }{\frac{d}{Q^3
d Q}
\Big( C_2(Q^2) \Big)  \Big|_{Q^2=0}} = \dfrac{1}{\beta^2}\,,
\end{align}
\textcolor{black}{which reproduces the analytic result
obtained in eq.(10)}.

\subsection{Gaussian-Dipole scenario}

Here we consider the photon distribution of Eq. (3)
for $\alpha=1$
(this choice is driven by the fact that in this case the procedure requires a
numerical integration) while for the proton one we consider the
Fourier Transform of the eff described by the dipole profile function of Ref.
[15]:

\begin{align}
    \label{strik}
      F_2^p(k_\perp) =
\left(1+\dfrac{k_\perp^2}{1.1 \, \mbox{GeV}^2}    \right)^{-2}~.
\end{align}

We change here the operator
\textcolor{black}{to
$\mathcal{O}^2_{Q^2}= d^4/(d Q^4)  |_{Q^2=0}$}
in order to remark that its choice can be optimized depending on the
mathematical difficulties in the procedure. Indeed the final
value of the extracted  $\langle b^n \rangle_p$ should not depend on its
particular choice.
distance should not.

With these settings,  { the right hand side of the equivalent of eq. (7)
becomes}:
\begin{align}
\frac{d^4}{ d Q^4} \Big( C^\gamma_2(Q^2) \Big)  \Big|_{Q^2=0}
&=-\frac{24}{\pi  }\,.
\end{align}
\textcolor{black}{At this point we should}
estimate the \textcolor{black}{corresponding} $\sigma_{eff}^{\gamma p}(Q^2)$. In
this case, an analytic
expression for this quantity {can not be obtained}. Therefore we numerically
evaluate $\sigma_{eff}^{\gamma
p}(Q^2)$ and then find a good fitting function of it.
 In particular, since the {chosen} operator requires {to take} the limit  $Q^2
\rightarrow 0$, it is {mandatory} to provide a realistic fitting function of
$[\sigma_{eff}^{\gamma p}(Q^2)]^{-1}$ in this region. For example, as shown
in Fig. \ref{fit}, a good functional form is:
\begin{align}
 \Big[\sigma_{eff}^{\gamma
p~}(Q^2)\Big]^{-1} \sim 14.7185 Q^6-8.18619 Q^5-2.315 Q^4+1.42751 Q^2+0.0533711
Q~.
\end{align}
{where it is understood that the numerical coefficients
have the correct dimension to reproduce the dimension on the left hand side}.
Within this choice:

\begin{align}
 \frac{d^4}{ d Q^4}   \Big[\sigma_{eff}^{\gamma
p~}(Q^2)\Big]^{-1}  \Big|_{Q^2=0} =  -55.56~\mbox{GeV}^{-2}.
\end{align}
By comparing the eq. (16) to eq. (18) we get:
 ${\langle b_\perp^2 \rangle_p} \sim 7.2728 $ GeV$^{-2}$.
The value  extracted from the proton  eff eq. (15)
 following the procedure presented in Refs. [31-32]
is given by
\begin{align}
 \langle b_\perp^2 \rangle_p = -\dfrac{d}{k_\perp d k_\perp}F_2^p (k_\perp)
\Big|_{k_\perp=0}  = 7.2727 ~\mbox{GeV}^{-2}.
\end{align}
{Those two values are rather close
and this fact let us positively conclude on the effectiveness of the proposed
procedure.}

operator.

 \begin{figure*}[t]
 \begin{center}
\includegraphics[scale=1]{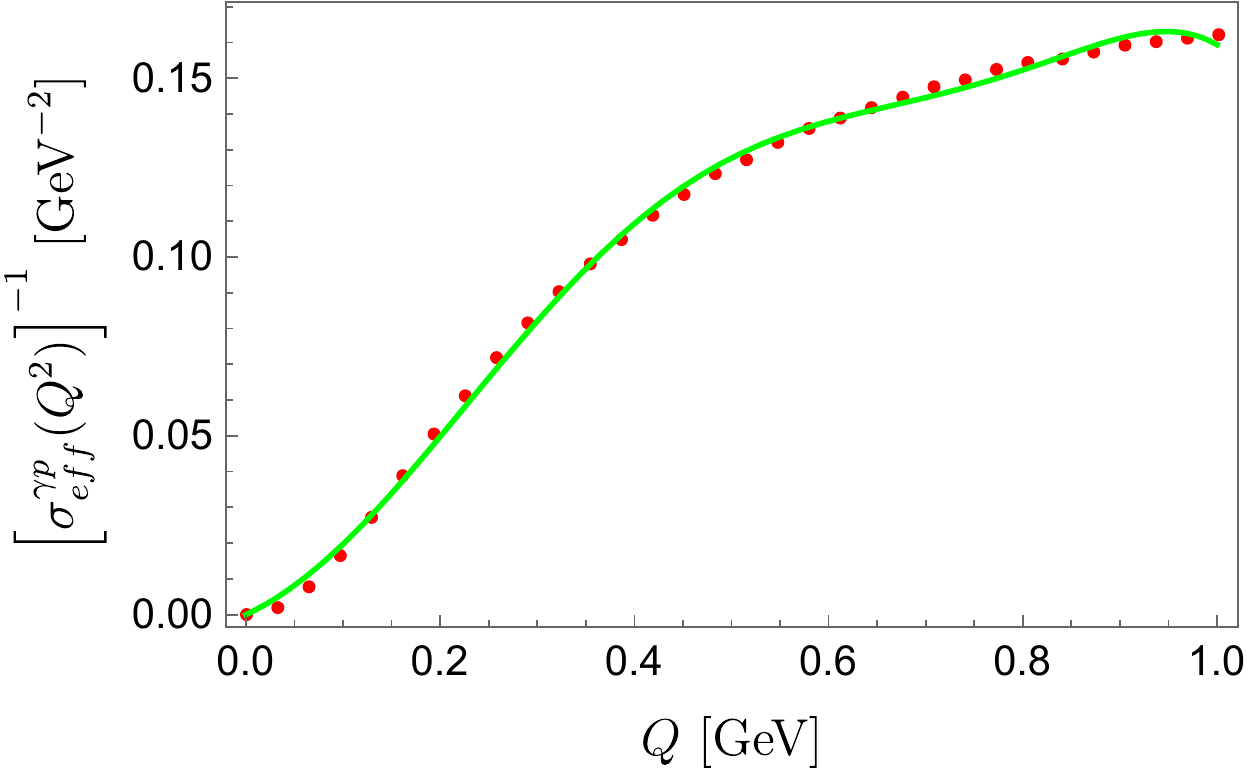}
\caption{The calculation of $[ \sigma_{eff}^{\gamma p}(Q^2 ]^{-1}$
within the eff eq. (15) and photon Gaussian distribution eq.(3) for $\alpha=1$.
Dots represent the numerical calculation and the full line is the fitting
function
eq. (17).}
\label{fit}
\end{center}
\end{figure*}

\bibliographystyle{apsrev4-1}
\bibliography{glasgow_mpi.bib}

\begin{thebibliography}{62}%
\makeatletter
\providecommand \@ifxundefined [1]{%
 \@ifx{#1\undefined}
}%
\providecommand \@ifnum [1]{%
 \ifnum #1\expandafter \@firstoftwo
 \else \expandafter \@secondoftwo
 \fi
}%
\providecommand \@ifx [1]{%
 \ifx #1\expandafter \@firstoftwo
 \else \expandafter \@secondoftwo
 \fi
}%
\providecommand \natexlab [1]{#1}%
\providecommand \enquote  [1]{``#1''}%
\providecommand \bibnamefont  [1]{#1}%
\providecommand \bibfnamefont [1]{#1}%
\providecommand \citenamefont [1]{#1}%
\providecommand \href@noop [0]{\@secondoftwo}%
\providecommand \href [0]{\begingroup \@sanitize@url \@href}%
\providecommand \@href[1]{\@@startlink{#1}\@@href}%
\providecommand \@@href[1]{\endgroup#1\@@endlink}%
\providecommand \@sanitize@url [0]{\catcode `\\12\catcode `\$12\catcode
  `\&12\catcode `\#12\catcode `\^12\catcode `\_12\catcode `\%12\relax}%
\providecommand \@@startlink[1]{}%
\providecommand \@@endlink[0]{}%
\providecommand \url  [0]{\begingroup\@sanitize@url \@url }%
\providecommand \@url [1]{\endgroup\@href {#1}{\urlprefix }}%
\providecommand \urlprefix  [0]{URL }%
\providecommand \Eprint [0]{\href }%
\providecommand \doibase [0]{http://dx.doi.org/}%
\providecommand \selectlanguage [0]{\@gobble}%
\providecommand \bibinfo  [0]{\@secondoftwo}%
\providecommand \bibfield  [0]{\@secondoftwo}%
\providecommand \translation [1]{[#1]}%
\providecommand \BibitemOpen [0]{}%
\providecommand \bibitemStop [0]{}%
\providecommand \bibitemNoStop [0]{.\EOS\space}%
\providecommand \EOS [0]{\spacefactor3000\relax}%
\providecommand \BibitemShut  [1]{\csname bibitem#1\endcsname}%
\let\auto@bib@innerbib\@empty
\bibitem [{\citenamefont {Goebel}\ \emph {et~al.}(1980)\citenamefont {Goebel},
  \citenamefont {Halzen},\ and\ \citenamefont {Scott}}]{Goebel:1979mi}%
  \BibitemOpen
  \bibfield  {author} {\bibinfo {author} {\bibfnamefont {C.}~\bibnamefont
  {Goebel}}, \bibinfo {author} {\bibfnamefont {F.}~\bibnamefont {Halzen}}, \
  and\ \bibinfo {author} {\bibfnamefont {D.~M.}\ \bibnamefont {Scott}},\ }\href
  {\doibase 10.1103/PhysRevD.22.2789} {\bibfield  {journal} {\bibinfo
  {journal} {Phys. Rev. D}\ }\textbf {\bibinfo {volume} {22}},\ \bibinfo
  {pages} {2789} (\bibinfo {year} {1980})}\BibitemShut {NoStop}%
\bibitem [{\citenamefont {Humpert}(1983)}]{Humpert:1983pw}%
  \BibitemOpen
  \bibfield  {author} {\bibinfo {author} {\bibfnamefont {B.}~\bibnamefont
  {Humpert}},\ }\href {\doibase 10.1016/0370-2693(83)90540-3} {\bibfield
  {journal} {\bibinfo  {journal} {Phys. Lett. B}\ }\textbf {\bibinfo {volume}
  {131}},\ \bibinfo {pages} {461} (\bibinfo {year} {1983})}\BibitemShut
  {NoStop}%
\bibitem [{\citenamefont {Mekhfi}(1985{\natexlab{a}})}]{Mekhfi:1983az}%
  \BibitemOpen
  \bibfield  {author} {\bibinfo {author} {\bibfnamefont {M.}~\bibnamefont
  {Mekhfi}},\ }\href {\doibase 10.1103/PhysRevD.32.2371} {\bibfield  {journal}
  {\bibinfo  {journal} {Phys. Rev. D}\ }\textbf {\bibinfo {volume} {32}},\
  \bibinfo {pages} {2371} (\bibinfo {year} {1985}{\natexlab{a}})}\BibitemShut
  {NoStop}%
\bibitem [{\citenamefont {Mekhfi}(1985{\natexlab{b}})}]{Mekhfi:1985dv}%
  \BibitemOpen
  \bibfield  {author} {\bibinfo {author} {\bibfnamefont {M.}~\bibnamefont
  {Mekhfi}},\ }\href {\doibase 10.1103/PhysRevD.32.2380} {\bibfield  {journal}
  {\bibinfo  {journal} {Phys. Rev. D}\ }\textbf {\bibinfo {volume} {32}},\
  \bibinfo {pages} {2380} (\bibinfo {year} {1985}{\natexlab{b}})}\BibitemShut
  {NoStop}%
\bibitem [{\citenamefont {Humpert}\ and\ \citenamefont
  {Odorico}(1985)}]{Humpert:1984ay}%
  \BibitemOpen
  \bibfield  {author} {\bibinfo {author} {\bibfnamefont {B.}~\bibnamefont
  {Humpert}}\ and\ \bibinfo {author} {\bibfnamefont {R.}~\bibnamefont
  {Odorico}},\ }\href {\doibase 10.1016/0370-2693(85)90587-8} {\bibfield
  {journal} {\bibinfo  {journal} {Phys. Lett. B}\ }\textbf {\bibinfo {volume}
  {154}},\ \bibinfo {pages} {211} (\bibinfo {year} {1985})}\BibitemShut
  {NoStop}%
\bibitem [{\citenamefont {Mangano}(1989)}]{Mangano:1988sq}%
  \BibitemOpen
  \bibfield  {author} {\bibinfo {author} {\bibfnamefont {M.~L.}\ \bibnamefont
  {Mangano}},\ }\href {\doibase 10.1007/BF01555875} {\bibfield  {journal}
  {\bibinfo  {journal} {Z. Phys. C}\ }\textbf {\bibinfo {volume} {42}},\
  \bibinfo {pages} {331} (\bibinfo {year} {1989})}\BibitemShut {NoStop}%
\bibitem [{\citenamefont {Paver}\ and\ \citenamefont {Treleani}(1982)}]{1a}%
  \BibitemOpen
  \bibfield  {author} {\bibinfo {author} {\bibfnamefont {N.}~\bibnamefont
  {Paver}}\ and\ \bibinfo {author} {\bibfnamefont {D.}~\bibnamefont
  {Treleani}},\ }\href {\doibase 10.1007/BF02814035} {\bibfield  {journal}
  {\bibinfo  {journal} {Nuovo Cim. A}\ }\textbf {\bibinfo {volume} {70}},\
  \bibinfo {pages} {215} (\bibinfo {year} {1982})}\BibitemShut {NoStop}%
\bibitem [{\citenamefont {Sjostrand}\ and\ \citenamefont {van
  Zijl}(1987)}]{Sjostrand:1986ep}%
  \BibitemOpen
  \bibfield  {author} {\bibinfo {author} {\bibfnamefont {T.}~\bibnamefont
  {Sjostrand}}\ and\ \bibinfo {author} {\bibfnamefont {M.}~\bibnamefont {van
  Zijl}},\ }\href {\doibase 10.1016/0370-2693(87)90722-2} {\bibfield  {journal}
  {\bibinfo  {journal} {Phys. Lett.}\ }\textbf {\bibinfo {volume} {B188}},\
  \bibinfo {pages} {149} (\bibinfo {year} {1987})}\BibitemShut {NoStop}%
\bibitem [{\citenamefont {Diehl}\ \emph {et~al.}(2012)\citenamefont {Diehl},
  \citenamefont {Ostermeier},\ and\ \citenamefont {Schafer}}]{4a}%
  \BibitemOpen
  \bibfield  {author} {\bibinfo {author} {\bibfnamefont {M.}~\bibnamefont
  {Diehl}}, \bibinfo {author} {\bibfnamefont {D.}~\bibnamefont {Ostermeier}}, \
  and\ \bibinfo {author} {\bibfnamefont {A.}~\bibnamefont {Schafer}},\ }\href
  {\doibase 10.1007/JHEP03(2012)089; 10.1007/JHEP03(2016)001} {\bibfield
  {journal} {\bibinfo  {journal} {JHEP}\ }\textbf {\bibinfo {volume} {03}},\
  \bibinfo {pages} {089} (\bibinfo {year} {2012})},\ \bibinfo {note} {[Erratum:
  JHEP03,001(2016)]},\ \Eprint {http://arxiv.org/abs/1111.0910}
  {arXiv:1111.0910 [hep-ph]} \BibitemShut {NoStop}%
\bibitem [{\citenamefont {Diehl}\ and\ \citenamefont
  {Schafer}(2011)}]{Diehl:2011tt}%
  \BibitemOpen
  \bibfield  {author} {\bibinfo {author} {\bibfnamefont {M.}~\bibnamefont
  {Diehl}}\ and\ \bibinfo {author} {\bibfnamefont {A.}~\bibnamefont
  {Schafer}},\ }\href {\doibase 10.1016/j.physletb.2011.03.024} {\bibfield
  {journal} {\bibinfo  {journal} {Phys. Lett.}\ }\textbf {\bibinfo {volume}
  {B698}},\ \bibinfo {pages} {389} (\bibinfo {year} {2011})},\ \Eprint
  {http://arxiv.org/abs/1102.3081} {arXiv:1102.3081 [hep-ph]} \BibitemShut
  {NoStop}%
\bibitem [{\citenamefont {Manohar}\ and\ \citenamefont {Waalewijn}(2012)}]{5a}%
  \BibitemOpen
  \bibfield  {author} {\bibinfo {author} {\bibfnamefont {A.~V.}\ \bibnamefont
  {Manohar}}\ and\ \bibinfo {author} {\bibfnamefont {W.~J.}\ \bibnamefont
  {Waalewijn}},\ }\href {\doibase 10.1103/PhysRevD.85.114009} {\bibfield
  {journal} {\bibinfo  {journal} {Phys. Rev.}\ }\textbf {\bibinfo {volume}
  {85}},\ \bibinfo {pages} {114009} (\bibinfo {year} {2012})}\BibitemShut
  {NoStop}%
\bibitem [{\citenamefont {Rinaldi}\ \emph {et~al.}(2013)\citenamefont
  {Rinaldi}, \citenamefont {Scopetta},\ and\ \citenamefont {Vento}}]{noi1}%
  \BibitemOpen
  \bibfield  {author} {\bibinfo {author} {\bibfnamefont {M.}~\bibnamefont
  {Rinaldi}}, \bibinfo {author} {\bibfnamefont {S.}~\bibnamefont {Scopetta}}, \
  and\ \bibinfo {author} {\bibfnamefont {V.}~\bibnamefont {Vento}},\ }\href
  {\doibase 10.1103/PhysRevD.87.114021} {\bibfield  {journal} {\bibinfo
  {journal} {Phys. Rev.}\ }\textbf {\bibinfo {volume} {D87}},\ \bibinfo {pages}
  {114021} (\bibinfo {year} {2013})},\ \Eprint {http://arxiv.org/abs/1302.6462}
  {arXiv:1302.6462 [hep-ph]} \BibitemShut {NoStop}%
\bibitem [{\citenamefont {Blok}\ \emph {et~al.}(2012)\citenamefont {Blok},
  \citenamefont {Dokshitser}, \citenamefont {Frankfurt},\ and\ \citenamefont
  {Strikman}}]{blok_1}%
  \BibitemOpen
  \bibfield  {author} {\bibinfo {author} {\bibfnamefont {B.}~\bibnamefont
  {Blok}}, \bibinfo {author} {\bibfnamefont {{\relax Yu}.}~\bibnamefont
  {Dokshitser}}, \bibinfo {author} {\bibfnamefont {L.}~\bibnamefont
  {Frankfurt}}, \ and\ \bibinfo {author} {\bibfnamefont {M.}~\bibnamefont
  {Strikman}},\ }\href {\doibase 10.1140/epjc/s10052-012-1963-8} {\bibfield
  {journal} {\bibinfo  {journal} {Eur. Phys. J.}\ }\textbf {\bibinfo {volume}
  {C72}},\ \bibinfo {pages} {1963} (\bibinfo {year} {2012})},\ \Eprint
  {http://arxiv.org/abs/1106.5533} {arXiv:1106.5533 [hep-ph]} \BibitemShut
  {NoStop}%
\bibitem [{\citenamefont {Blok}\ \emph {et~al.}(2011)\citenamefont {Blok},
  \citenamefont {Dokshitzer}, \citenamefont {Frankfurt},\ and\ \citenamefont
  {Strikman}}]{Blok1}%
  \BibitemOpen
  \bibfield  {author} {\bibinfo {author} {\bibfnamefont {B.}~\bibnamefont
  {Blok}}, \bibinfo {author} {\bibfnamefont {{\relax Yu}.}~\bibnamefont
  {Dokshitzer}}, \bibinfo {author} {\bibfnamefont {L.}~\bibnamefont
  {Frankfurt}}, \ and\ \bibinfo {author} {\bibfnamefont {M.}~\bibnamefont
  {Strikman}},\ }\href {\doibase 10.1103/PhysRevD.83.071501} {\bibfield
  {journal} {\bibinfo  {journal} {Phys. Rev.}\ }\textbf {\bibinfo {volume}
  {D83}},\ \bibinfo {pages} {071501} (\bibinfo {year} {2011})},\ \Eprint
  {http://arxiv.org/abs/1009.2714} {arXiv:1009.2714 [hep-ph]} \BibitemShut
  {NoStop}%
\bibitem [{\citenamefont {Blok}\ \emph {et~al.}(2014)\citenamefont {Blok},
  \citenamefont {Dokshitzer}, \citenamefont {Frankfurt},\ and\ \citenamefont
  {Strikman}}]{Blok2}%
  \BibitemOpen
  \bibfield  {author} {\bibinfo {author} {\bibfnamefont {B.}~\bibnamefont
  {Blok}}, \bibinfo {author} {\bibfnamefont {{\relax Yu}.}~\bibnamefont
  {Dokshitzer}}, \bibinfo {author} {\bibfnamefont {L.}~\bibnamefont
  {Frankfurt}}, \ and\ \bibinfo {author} {\bibfnamefont {M.}~\bibnamefont
  {Strikman}},\ }\href {\doibase 10.1140/epjc/s10052-014-2926-z} {\bibfield
  {journal} {\bibinfo  {journal} {Eur. Phys. J.}\ }\textbf {\bibinfo {volume}
  {C74}},\ \bibinfo {pages} {2926} (\bibinfo {year} {2014})},\ \Eprint
  {http://arxiv.org/abs/1306.3763} {arXiv:1306.3763 [hep-ph]} \BibitemShut
  {NoStop}%
\bibitem [{\citenamefont {Gaunt}\ and\ \citenamefont
  {Stirling}(2010)}]{gauntevo}%
  \BibitemOpen
  \bibfield  {author} {\bibinfo {author} {\bibfnamefont {J.~R.}\ \bibnamefont
  {Gaunt}}\ and\ \bibinfo {author} {\bibfnamefont {W.~J.}\ \bibnamefont
  {Stirling}},\ }\href {\doibase 10.1007/JHEP03(2010)005} {\bibfield  {journal}
  {\bibinfo  {journal} {JHEP}\ ,\ \bibinfo {pages} {005}} (\bibinfo {year}
  {2010})},\ \Eprint {http://arxiv.org/abs/0910.4347} {arXiv:0910.4347
  [hep-ph]} \BibitemShut {NoStop}%
\bibitem [{\citenamefont {Chang}\ \emph {et~al.}(2013)\citenamefont {Chang},
  \citenamefont {Manohar},\ and\ \citenamefont {Waalewijn}}]{man}%
  \BibitemOpen
  \bibfield  {author} {\bibinfo {author} {\bibfnamefont {H.-M.}\ \bibnamefont
  {Chang}}, \bibinfo {author} {\bibfnamefont {A.~V.}\ \bibnamefont {Manohar}},
  \ and\ \bibinfo {author} {\bibfnamefont {W.~J.}\ \bibnamefont {Waalewijn}},\
  }\href {\doibase 10.1103/PhysRevD.87.034009} {\bibfield  {journal} {\bibinfo
  {journal} {Phys. Rev.}\ }\textbf {\bibinfo {volume} {D87}},\ \bibinfo {pages}
  {034009} (\bibinfo {year} {2013})},\ \Eprint {http://arxiv.org/abs/1211.3132}
  {arXiv:1211.3132 [hep-ph]} \BibitemShut {NoStop}%
\bibitem [{\citenamefont {Rinaldi}\ \emph
  {et~al.}(2016{\natexlab{a}})\citenamefont {Rinaldi}, \citenamefont
  {Scopetta}, \citenamefont {Traini},\ and\ \citenamefont {Vento}}]{noij2}%
  \BibitemOpen
  \bibfield  {author} {\bibinfo {author} {\bibfnamefont {M.}~\bibnamefont
  {Rinaldi}}, \bibinfo {author} {\bibfnamefont {S.}~\bibnamefont {Scopetta}},
  \bibinfo {author} {\bibfnamefont {M.~C.}\ \bibnamefont {Traini}}, \ and\
  \bibinfo {author} {\bibfnamefont {V.}~\bibnamefont {Vento}},\ }\href
  {\doibase 10.1007/JHEP10(2016)063} {\bibfield  {journal} {\bibinfo  {journal}
  {JHEP}\ }\textbf {\bibinfo {volume} {10}},\ \bibinfo {pages} {063} (\bibinfo
  {year} {2016}{\natexlab{a}})},\ \Eprint {http://arxiv.org/abs/1608.02521}
  {arXiv:1608.02521 [hep-ph]} \BibitemShut {NoStop}%
\bibitem [{\citenamefont {Rinaldi}\ \emph {et~al.}(2014)\citenamefont
  {Rinaldi}, \citenamefont {Scopetta}, \citenamefont {Traini},\ and\
  \citenamefont {Vento}}]{noij1}%
  \BibitemOpen
  \bibfield  {author} {\bibinfo {author} {\bibfnamefont {M.}~\bibnamefont
  {Rinaldi}}, \bibinfo {author} {\bibfnamefont {S.}~\bibnamefont {Scopetta}},
  \bibinfo {author} {\bibfnamefont {M.}~\bibnamefont {Traini}}, \ and\ \bibinfo
  {author} {\bibfnamefont {V.}~\bibnamefont {Vento}},\ }\href {\doibase
  10.1007/JHEP12(2014)028} {\bibfield  {journal} {\bibinfo  {journal} {JHEP}\
  }\textbf {\bibinfo {volume} {12}},\ \bibinfo {pages} {028} (\bibinfo {year}
  {2014})},\ \Eprint {http://arxiv.org/abs/1409.1500} {arXiv:1409.1500
  [hep-ph]} \BibitemShut {NoStop}%
\bibitem [{\citenamefont {Diehl}\ \emph {et~al.}(2020)\citenamefont {Diehl},
  \citenamefont {Gaunt}, \citenamefont {Lang}, \citenamefont {Pl\"o\ss{}l},\
  and\ \citenamefont {Sch\"afer}}]{Diehl:2020xyg}%
  \BibitemOpen
  \bibfield  {author} {\bibinfo {author} {\bibfnamefont {M.}~\bibnamefont
  {Diehl}}, \bibinfo {author} {\bibfnamefont {J.~R.}\ \bibnamefont {Gaunt}},
  \bibinfo {author} {\bibfnamefont {D.~M.}\ \bibnamefont {Lang}}, \bibinfo
  {author} {\bibfnamefont {P.}~\bibnamefont {Pl\"o\ss{}l}}, \ and\ \bibinfo
  {author} {\bibfnamefont {A.}~\bibnamefont {Sch\"afer}},\ }\href {\doibase
  10.1140/epjc/s10052-020-8038-z} {\bibfield  {journal} {\bibinfo  {journal}
  {Eur. Phys. J. C}\ }\textbf {\bibinfo {volume} {80}},\ \bibinfo {pages} {468}
  (\bibinfo {year} {2020})},\ \Eprint {http://arxiv.org/abs/2001.10428}
  {arXiv:2001.10428 [hep-ph]} \BibitemShut {NoStop}%
\bibitem [{\citenamefont {Diehl}\ \emph
  {et~al.}(2019{\natexlab{a}})\citenamefont {Diehl}, \citenamefont
  {Pl\"o\ss{}l},\ and\ \citenamefont {Sch\"afer}}]{Diehl:2018kgr}%
  \BibitemOpen
  \bibfield  {author} {\bibinfo {author} {\bibfnamefont {M.}~\bibnamefont
  {Diehl}}, \bibinfo {author} {\bibfnamefont {P.}~\bibnamefont {Pl\"o\ss{}l}},
  \ and\ \bibinfo {author} {\bibfnamefont {A.}~\bibnamefont {Sch\"afer}},\
  }\href {\doibase 10.1140/epjc/s10052-019-6777-5} {\bibfield  {journal}
  {\bibinfo  {journal} {Eur. Phys. J. C}\ }\textbf {\bibinfo {volume} {79}},\
  \bibinfo {pages} {253} (\bibinfo {year} {2019}{\natexlab{a}})},\ \Eprint
  {http://arxiv.org/abs/1811.00289} {arXiv:1811.00289 [hep-ph]} \BibitemShut
  {NoStop}%
\bibitem [{\citenamefont {Diehl}\ \emph {et~al.}(2017)\citenamefont {Diehl},
  \citenamefont {Gaunt},\ and\ \citenamefont {Sch\"onwald}}]{Diehl:2017kgu}%
  \BibitemOpen
  \bibfield  {author} {\bibinfo {author} {\bibfnamefont {M.}~\bibnamefont
  {Diehl}}, \bibinfo {author} {\bibfnamefont {J.~R.}\ \bibnamefont {Gaunt}}, \
  and\ \bibinfo {author} {\bibfnamefont {K.}~\bibnamefont {Sch\"onwald}},\
  }\href {\doibase 10.1007/JHEP06(2017)083} {\bibfield  {journal} {\bibinfo
  {journal} {JHEP}\ }\textbf {\bibinfo {volume} {06}},\ \bibinfo {pages} {083}
  (\bibinfo {year} {2017})},\ \Eprint {http://arxiv.org/abs/1702.06486}
  {arXiv:1702.06486 [hep-ph]} \BibitemShut {NoStop}%
\bibitem [{\citenamefont {Diehl}\ \emph {et~al.}(2016)\citenamefont {Diehl},
  \citenamefont {Gaunt}, \citenamefont {Ostermeier}, \citenamefont
  {Pl\"o\ss{}l},\ and\ \citenamefont {Sch\"afer}}]{Diehl:2015bca}%
  \BibitemOpen
  \bibfield  {author} {\bibinfo {author} {\bibfnamefont {M.}~\bibnamefont
  {Diehl}}, \bibinfo {author} {\bibfnamefont {J.~R.}\ \bibnamefont {Gaunt}},
  \bibinfo {author} {\bibfnamefont {D.}~\bibnamefont {Ostermeier}}, \bibinfo
  {author} {\bibfnamefont {P.}~\bibnamefont {Pl\"o\ss{}l}}, \ and\ \bibinfo
  {author} {\bibfnamefont {A.}~\bibnamefont {Sch\"afer}},\ }\href {\doibase
  10.1007/JHEP01(2016)076} {\bibfield  {journal} {\bibinfo  {journal} {JHEP}\
  }\textbf {\bibinfo {volume} {01}},\ \bibinfo {pages} {076} (\bibinfo {year}
  {2016})},\ \Eprint {http://arxiv.org/abs/1510.08696} {arXiv:1510.08696
  [hep-ph]} \BibitemShut {NoStop}%
\bibitem [{\citenamefont {Gaunt}(2014)}]{Gaunt:2014ska}%
  \BibitemOpen
  \bibfield  {author} {\bibinfo {author} {\bibfnamefont {J.~R.}\ \bibnamefont
  {Gaunt}},\ }\href {\doibase 10.1007/JHEP07(2014)110} {\bibfield  {journal}
  {\bibinfo  {journal} {JHEP}\ }\textbf {\bibinfo {volume} {07}},\ \bibinfo
  {pages} {110} (\bibinfo {year} {2014})},\ \Eprint
  {http://arxiv.org/abs/1405.2080} {arXiv:1405.2080 [hep-ph]} \BibitemShut
  {NoStop}%
\bibitem [{\citenamefont {Diehl}\ \emph
  {et~al.}(2019{\natexlab{b}})\citenamefont {Diehl}, \citenamefont {Gaunt},
  \citenamefont {Pl\"o\ss{}l},\ and\ \citenamefont
  {Sch\"afer}}]{Diehl:2019rdh}%
  \BibitemOpen
  \bibfield  {author} {\bibinfo {author} {\bibfnamefont {M.}~\bibnamefont
  {Diehl}}, \bibinfo {author} {\bibfnamefont {J.~R.}\ \bibnamefont {Gaunt}},
  \bibinfo {author} {\bibfnamefont {P.}~\bibnamefont {Pl\"o\ss{}l}}, \ and\
  \bibinfo {author} {\bibfnamefont {A.}~\bibnamefont {Sch\"afer}},\ }\href
  {\doibase 10.21468/SciPostPhys.7.2.017} {\bibfield  {journal} {\bibinfo
  {journal} {SciPost Phys.}\ }\textbf {\bibinfo {volume} {7}},\ \bibinfo
  {pages} {017} (\bibinfo {year} {2019}{\natexlab{b}})},\ \Eprint
  {http://arxiv.org/abs/1902.08019} {arXiv:1902.08019 [hep-ph]} \BibitemShut
  {NoStop}%
\bibitem [{\citenamefont {Diehl}\ and\ \citenamefont
  {Nagar}(2019)}]{Diehl:2018wfy}%
  \BibitemOpen
  \bibfield  {author} {\bibinfo {author} {\bibfnamefont {M.}~\bibnamefont
  {Diehl}}\ and\ \bibinfo {author} {\bibfnamefont {R.}~\bibnamefont {Nagar}},\
  }\href {\doibase 10.1007/JHEP04(2019)124} {\bibfield  {journal} {\bibinfo
  {journal} {JHEP}\ }\textbf {\bibinfo {volume} {04}},\ \bibinfo {pages} {124}
  (\bibinfo {year} {2019})},\ \Eprint {http://arxiv.org/abs/1812.09509}
  {arXiv:1812.09509 [hep-ph]} \BibitemShut {NoStop}%
\bibitem [{\citenamefont {Ryskin}\ and\ \citenamefont
  {Snigirev}(2011)}]{Ryskin:2011kk}%
  \BibitemOpen
  \bibfield  {author} {\bibinfo {author} {\bibfnamefont {M.~G.}\ \bibnamefont
  {Ryskin}}\ and\ \bibinfo {author} {\bibfnamefont {A.~M.}\ \bibnamefont
  {Snigirev}},\ }\href {\doibase 10.1103/PhysRevD.83.114047} {\bibfield
  {journal} {\bibinfo  {journal} {Phys. Rev.}\ }\textbf {\bibinfo {volume}
  {D83}},\ \bibinfo {pages} {114047} (\bibinfo {year} {2011})},\ \Eprint
  {http://arxiv.org/abs/1103.3495} {arXiv:1103.3495 [hep-ph]} \BibitemShut
  {NoStop}%
\bibitem [{\citenamefont {Ryskin}\ and\ \citenamefont
  {Snigirev}(2012)}]{Ryskin:2012qx}%
  \BibitemOpen
  \bibfield  {author} {\bibinfo {author} {\bibfnamefont {M.~G.}\ \bibnamefont
  {Ryskin}}\ and\ \bibinfo {author} {\bibfnamefont {A.~M.}\ \bibnamefont
  {Snigirev}},\ }\href {\doibase 10.1103/PhysRevD.86.014018} {\bibfield
  {journal} {\bibinfo  {journal} {Phys. Rev.}\ }\textbf {\bibinfo {volume}
  {D86}},\ \bibinfo {pages} {014018} (\bibinfo {year} {2012})},\ \Eprint
  {http://arxiv.org/abs/1203.2330} {arXiv:1203.2330 [hep-ph]} \BibitemShut
  {NoStop}%
\bibitem [{\citenamefont {Calucci}\ and\ \citenamefont
  {Treleani}(1999)}]{Calucci:1999yz}%
  \BibitemOpen
  \bibfield  {author} {\bibinfo {author} {\bibfnamefont {G.}~\bibnamefont
  {Calucci}}\ and\ \bibinfo {author} {\bibfnamefont {D.}~\bibnamefont
  {Treleani}},\ }\href {\doibase 10.1103/PhysRevD.60.054023} {\bibfield
  {journal} {\bibinfo  {journal} {Phys. Rev.}\ }\textbf {\bibinfo {volume}
  {D60}},\ \bibinfo {pages} {054023} (\bibinfo {year} {1999})},\ \Eprint
  {http://arxiv.org/abs/hep-ph/9902479} {arXiv:hep-ph/9902479 [hep-ph]}
  \BibitemShut {NoStop}%
\bibitem [{\citenamefont {Aaij}\ \emph {et~al.}(2020)\citenamefont {Aaij} \emph
  {et~al.}}]{Aaij:2020smi}%
  \BibitemOpen
  \bibfield  {author} {\bibinfo {author} {\bibfnamefont {R.}~\bibnamefont
  {Aaij}} \emph {et~al.} (\bibinfo {collaboration} {LHCb}),\ }\href {\doibase
  10.1103/PhysRevLett.125.212001} {\bibfield  {journal} {\bibinfo  {journal}
  {Phys. Rev. Lett.}\ }\textbf {\bibinfo {volume} {125}},\ \bibinfo {pages}
  {212001} (\bibinfo {year} {2020})},\ \Eprint
  {http://arxiv.org/abs/2007.06945} {arXiv:2007.06945 [hep-ex]} \BibitemShut
  {NoStop}%
\bibitem [{\citenamefont {Rinaldi}\ and\ \citenamefont
  {Ceccopieri}(2018)}]{rapid}%
  \BibitemOpen
  \bibfield  {author} {\bibinfo {author} {\bibfnamefont {M.}~\bibnamefont
  {Rinaldi}}\ and\ \bibinfo {author} {\bibfnamefont {F.~A.}\ \bibnamefont
  {Ceccopieri}},\ }\href {\doibase 10.1103/PhysRevD.97.071501} {\bibfield
  {journal} {\bibinfo  {journal} {Phys. Rev.}\ }\textbf {\bibinfo {volume}
  {D97}},\ \bibinfo {pages} {071501} (\bibinfo {year} {2018})},\ \Eprint
  {http://arxiv.org/abs/1801.04760} {arXiv:1801.04760 [hep-ph]} \BibitemShut
  {NoStop}%
\bibitem [{\citenamefont {Rinaldi}\ and\ \citenamefont
  {Ceccopieri}(2019)}]{jhepc}%
  \BibitemOpen
  \bibfield  {author} {\bibinfo {author} {\bibfnamefont {M.}~\bibnamefont
  {Rinaldi}}\ and\ \bibinfo {author} {\bibfnamefont {F.~A.}\ \bibnamefont
  {Ceccopieri}},\ }\href {\doibase 10.1007/JHEP09(2019)097} {\bibfield
  {journal} {\bibinfo  {journal} {JHEP}\ }\textbf {\bibinfo {volume} {09}},\
  \bibinfo {pages} {097} (\bibinfo {year} {2019})},\ \Eprint
  {http://arxiv.org/abs/1812.04286} {arXiv:1812.04286 [hep-ph]} \BibitemShut
  {NoStop}%
\bibitem [{\citenamefont {Aaboud}\ \emph {et~al.}(2016)\citenamefont {Aaboud}
  \emph {et~al.}}]{4jet}%
  \BibitemOpen
  \bibfield  {author} {\bibinfo {author} {\bibfnamefont {M.}~\bibnamefont
  {Aaboud}} \emph {et~al.} (\bibinfo {collaboration} {ATLAS}),\ }\href
  {\doibase 10.1007/JHEP11(2016)110} {\bibfield  {journal} {\bibinfo  {journal}
  {JHEP}\ }\textbf {\bibinfo {volume} {11}},\ \bibinfo {pages} {110} (\bibinfo
  {year} {2016})},\ \Eprint {http://arxiv.org/abs/1608.01857} {arXiv:1608.01857
  [hep-ex]} \BibitemShut {NoStop}%
\bibitem [{\citenamefont {CMS}(2021)}]{CMS:2021ijt}%
  \BibitemOpen
  \bibfield  {author} {\bibinfo {author} {\bibnamefont {CMS}} (\bibinfo
  {collaboration} {CMS}),\ }\href@noop {} {\bibfield  {journal} {\bibinfo
  {journal} {CMS-PAS-SMP-20-007}\ } (\bibinfo {year} {2021})}\BibitemShut
  {NoStop}%
\bibitem [{\citenamefont {Butterworth}\ \emph {et~al.}(1996)\citenamefont
  {Butterworth}, \citenamefont {Forshaw},\ and\ \citenamefont
  {Seymour}}]{Butterworth:1996zw}%
  \BibitemOpen
  \bibfield  {author} {\bibinfo {author} {\bibfnamefont {J.~M.}\ \bibnamefont
  {Butterworth}}, \bibinfo {author} {\bibfnamefont {J.~R.}\ \bibnamefont
  {Forshaw}}, \ and\ \bibinfo {author} {\bibfnamefont {M.~H.}\ \bibnamefont
  {Seymour}},\ }\href {\doibase 10.1007/s002880050286} {\bibfield  {journal}
  {\bibinfo  {journal} {Z. Phys. C}\ }\textbf {\bibinfo {volume} {72}},\
  \bibinfo {pages} {637} (\bibinfo {year} {1996})},\ \Eprint
  {http://arxiv.org/abs/hep-ph/9601371} {arXiv:hep-ph/9601371} \BibitemShut
  {NoStop}%
\bibitem [{\citenamefont {Blok}\ and\ \citenamefont
  {Strikman}(2014)}]{Blok:2014rza}%
  \BibitemOpen
  \bibfield  {author} {\bibinfo {author} {\bibfnamefont {B.}~\bibnamefont
  {Blok}}\ and\ \bibinfo {author} {\bibfnamefont {M.}~\bibnamefont
  {Strikman}},\ }\href {\doibase 10.1140/epjc/s10052-014-3214-7} {\bibfield
  {journal} {\bibinfo  {journal} {Eur. Phys. J. C}\ }\textbf {\bibinfo {volume}
  {74}},\ \bibinfo {pages} {3214} (\bibinfo {year} {2014})},\ \Eprint
  {http://arxiv.org/abs/1410.5064} {arXiv:1410.5064 [hep-ph]} \BibitemShut
  {NoStop}%
\bibitem [{\citenamefont {Klasen}(2002)}]{Klasen:2002xb}%
  \BibitemOpen
  \bibfield  {author} {\bibinfo {author} {\bibfnamefont {M.}~\bibnamefont
  {Klasen}},\ }\href {\doibase 10.1103/RevModPhys.74.1221} {\bibfield
  {journal} {\bibinfo  {journal} {Rev. Mod. Phys.}\ }\textbf {\bibinfo {volume}
  {74}},\ \bibinfo {pages} {1221} (\bibinfo {year} {2002})},\ \Eprint
  {http://arxiv.org/abs/hep-ph/0206169} {arXiv:hep-ph/0206169} \BibitemShut
  {NoStop}%
\bibitem [{\citenamefont {Frankfurt}\ and\ \citenamefont
  {Strikman}(1981)}]{Frankfurt:1981mk}%
  \BibitemOpen
  \bibfield  {author} {\bibinfo {author} {\bibfnamefont {L.~L.}\ \bibnamefont
  {Frankfurt}}\ and\ \bibinfo {author} {\bibfnamefont {M.~I.}\ \bibnamefont
  {Strikman}},\ }\href {\doibase 10.1016/0370-1573(81)90129-0} {\bibfield
  {journal} {\bibinfo  {journal} {Phys. Rept.}\ }\textbf {\bibinfo {volume}
  {76}},\ \bibinfo {pages} {215} (\bibinfo {year} {1981})}\BibitemShut
  {NoStop}%
\bibitem [{\citenamefont {Nikolaev}\ and\ \citenamefont
  {Zakharov}(1991)}]{Nikolaev:1990ja}%
  \BibitemOpen
  \bibfield  {author} {\bibinfo {author} {\bibfnamefont {N.~N.}\ \bibnamefont
  {Nikolaev}}\ and\ \bibinfo {author} {\bibfnamefont {B.~G.}\ \bibnamefont
  {Zakharov}},\ }\href {\doibase 10.1007/BF01483577} {\bibfield  {journal}
  {\bibinfo  {journal} {Z. Phys. C}\ }\textbf {\bibinfo {volume} {49}},\
  \bibinfo {pages} {607} (\bibinfo {year} {1991})}\BibitemShut {NoStop}%
\bibitem [{\citenamefont {Rinaldi}\ \emph {et~al.}(2018)\citenamefont
  {Rinaldi}, \citenamefont {Scopetta}, \citenamefont {Traini},\ and\
  \citenamefont {Vento}}]{noipion}%
  \BibitemOpen
  \bibfield  {author} {\bibinfo {author} {\bibfnamefont {M.}~\bibnamefont
  {Rinaldi}}, \bibinfo {author} {\bibfnamefont {S.}~\bibnamefont {Scopetta}},
  \bibinfo {author} {\bibfnamefont {M.}~\bibnamefont {Traini}}, \ and\ \bibinfo
  {author} {\bibfnamefont {V.}~\bibnamefont {Vento}},\ }\href {\doibase
  10.1140/epjc/s10052-018-6256-4} {\bibfield  {journal} {\bibinfo  {journal}
  {Eur. Phys. J.}\ }\textbf {\bibinfo {volume} {C78}},\ \bibinfo {pages} {781}
  (\bibinfo {year} {2018})}\BibitemShut {NoStop}%
\bibitem [{\citenamefont {Kasemets}\ and\ \citenamefont
  {Mukherjee}(2016)}]{Kasemets:2016nio}%
  \BibitemOpen
  \bibfield  {author} {\bibinfo {author} {\bibfnamefont {T.}~\bibnamefont
  {Kasemets}}\ and\ \bibinfo {author} {\bibfnamefont {A.}~\bibnamefont
  {Mukherjee}},\ }\href {\doibase 10.1103/PhysRevD.94.074029} {\bibfield
  {journal} {\bibinfo  {journal} {Phys. Rev.}\ }\textbf {\bibinfo {volume}
  {D94}},\ \bibinfo {pages} {074029} (\bibinfo {year} {2016})},\ \Eprint
  {http://arxiv.org/abs/1606.05686} {arXiv:1606.05686 [hep-ph]} \BibitemShut
  {NoStop}%
\bibitem [{\citenamefont {Rinaldi}(2020)}]{Rinaldi:2020ybv}%
  \BibitemOpen
  \bibfield  {author} {\bibinfo {author} {\bibfnamefont {M.}~\bibnamefont
  {Rinaldi}},\ }\href {\doibase 10.1140/epjc/s10052-020-8241-y} {\bibfield
  {journal} {\bibinfo  {journal} {Eur. Phys. J.}\ }\textbf {\bibinfo {volume}
  {C80}},\ \bibinfo {pages} {678} (\bibinfo {year} {2020})},\ \Eprint
  {http://arxiv.org/abs/2003.09400} {arXiv:2003.09400 [hep-ph]} \BibitemShut
  {NoStop}%
\bibitem [{\citenamefont {Rinaldi}\ \emph
  {et~al.}(2016{\natexlab{b}})\citenamefont {Rinaldi}, \citenamefont
  {Scopetta}, \citenamefont {Traini},\ and\ \citenamefont {Vento}}]{noiPLB}%
  \BibitemOpen
  \bibfield  {author} {\bibinfo {author} {\bibfnamefont {M.}~\bibnamefont
  {Rinaldi}}, \bibinfo {author} {\bibfnamefont {S.}~\bibnamefont {Scopetta}},
  \bibinfo {author} {\bibfnamefont {M.}~\bibnamefont {Traini}}, \ and\ \bibinfo
  {author} {\bibfnamefont {V.}~\bibnamefont {Vento}},\ }\href {\doibase
  10.1016/j.physletb.2015.11.031} {\bibfield  {journal} {\bibinfo  {journal}
  {Phys. Lett.}\ }\textbf {\bibinfo {volume} {B752}},\ \bibinfo {pages} {40}
  (\bibinfo {year} {2016}{\natexlab{b}})},\ \Eprint
  {http://arxiv.org/abs/1506.05742} {arXiv:1506.05742 [hep-ph]} \BibitemShut
  {NoStop}%
\bibitem [{\citenamefont {Dorokhov}\ \emph {et~al.}(2006)\citenamefont
  {Dorokhov}, \citenamefont {Broniowski},\ and\ \citenamefont
  {Ruiz~Arriola}}]{arriola}%
  \BibitemOpen
  \bibfield  {author} {\bibinfo {author} {\bibfnamefont {A.~E.}\ \bibnamefont
  {Dorokhov}}, \bibinfo {author} {\bibfnamefont {W.}~\bibnamefont
  {Broniowski}}, \ and\ \bibinfo {author} {\bibfnamefont {E.}~\bibnamefont
  {Ruiz~Arriola}},\ }\href {\doibase 10.1103/PhysRevD.74.054023} {\bibfield
  {journal} {\bibinfo  {journal} {Phys. Rev.}\ }\textbf {\bibinfo {volume}
  {D74}},\ \bibinfo {pages} {054023} (\bibinfo {year} {2006})},\ \Eprint
  {http://arxiv.org/abs/hep-ph/0607171} {arXiv:hep-ph/0607171 [hep-ph]}
  \BibitemShut {NoStop}%
\bibitem [{\citenamefont {Brodsky}\ \emph {et~al.}(1994)\citenamefont
  {Brodsky}, \citenamefont {Frankfurt}, \citenamefont {Gunion}, \citenamefont
  {Mueller},\ and\ \citenamefont {Strikman}}]{brodfrank}%
  \BibitemOpen
  \bibfield  {author} {\bibinfo {author} {\bibfnamefont {S.~J.}\ \bibnamefont
  {Brodsky}}, \bibinfo {author} {\bibfnamefont {L.}~\bibnamefont {Frankfurt}},
  \bibinfo {author} {\bibfnamefont {J.~F.}\ \bibnamefont {Gunion}}, \bibinfo
  {author} {\bibfnamefont {A.~H.}\ \bibnamefont {Mueller}}, \ and\ \bibinfo
  {author} {\bibfnamefont {M.}~\bibnamefont {Strikman}},\ }\href {\doibase
  10.1103/PhysRevD.50.3134} {\bibfield  {journal} {\bibinfo  {journal} {Phys.
  Rev.}\ }\textbf {\bibinfo {volume} {D50}},\ \bibinfo {pages} {3134} (\bibinfo
  {year} {1994})},\ \Eprint {http://arxiv.org/abs/hep-ph/9402283}
  {arXiv:hep-ph/9402283 [hep-ph]} \BibitemShut {NoStop}%
\bibitem [{\citenamefont {Gaunt}(2013)}]{Gaunt:2012dd}%
  \BibitemOpen
  \bibfield  {author} {\bibinfo {author} {\bibfnamefont {J.~R.}\ \bibnamefont
  {Gaunt}},\ }\href {\doibase 10.1007/JHEP01(2013)042} {\bibfield  {journal}
  {\bibinfo  {journal} {JHEP}\ }\textbf {\bibinfo {volume} {01}},\ \bibinfo
  {pages} {042} (\bibinfo {year} {2013})},\ \Eprint
  {http://arxiv.org/abs/1207.0480} {arXiv:1207.0480 [hep-ph]} \BibitemShut
  {NoStop}%
\bibitem [{\citenamefont {Chekanov}\ \emph {et~al.}(2008)\citenamefont
  {Chekanov} \emph {et~al.}}]{Chekanov:2007ab}%
  \BibitemOpen
  \bibfield  {author} {\bibinfo {author} {\bibfnamefont {S.}~\bibnamefont
  {Chekanov}} \emph {et~al.} (\bibinfo {collaboration} {ZEUS}),\ }\href
  {\doibase 10.1016/j.nuclphysb.2007.08.021} {\bibfield  {journal} {\bibinfo
  {journal} {Nucl. Phys. B}\ }\textbf {\bibinfo {volume} {792}},\ \bibinfo
  {pages} {1} (\bibinfo {year} {2008})},\ \Eprint
  {http://arxiv.org/abs/0707.3749} {arXiv:0707.3749 [hep-ex]} \BibitemShut
  {NoStop}%
\bibitem [{\citenamefont {Marchesini}\ \emph {et~al.}(1992)\citenamefont
  {Marchesini}, \citenamefont {Webber}, \citenamefont {Abbiendi}, \citenamefont
  {Knowles}, \citenamefont {Seymour},\ and\ \citenamefont
  {Stanco}}]{Marchesini:1991ch}%
  \BibitemOpen
  \bibfield  {author} {\bibinfo {author} {\bibfnamefont {G.}~\bibnamefont
  {Marchesini}}, \bibinfo {author} {\bibfnamefont {B.~R.}\ \bibnamefont
  {Webber}}, \bibinfo {author} {\bibfnamefont {G.}~\bibnamefont {Abbiendi}},
  \bibinfo {author} {\bibfnamefont {I.~G.}\ \bibnamefont {Knowles}}, \bibinfo
  {author} {\bibfnamefont {M.~H.}\ \bibnamefont {Seymour}}, \ and\ \bibinfo
  {author} {\bibfnamefont {L.}~\bibnamefont {Stanco}},\ }\href {\doibase
  10.1016/0010-4655(92)90055-4} {\bibfield  {journal} {\bibinfo  {journal}
  {Comput. Phys. Commun.}\ }\textbf {\bibinfo {volume} {67}},\ \bibinfo {pages}
  {465} (\bibinfo {year} {1992})}\BibitemShut {NoStop}%
\bibitem [{\citenamefont {Sjostrand}\ \emph {et~al.}(2001)\citenamefont
  {Sjostrand}, \citenamefont {Eden}, \citenamefont {Friberg}, \citenamefont
  {Lonnblad}, \citenamefont {Miu}, \citenamefont {Mrenna},\ and\ \citenamefont
  {Norrbin}}]{Sjostrand:2000wi}%
  \BibitemOpen
  \bibfield  {author} {\bibinfo {author} {\bibfnamefont {T.}~\bibnamefont
  {Sjostrand}}, \bibinfo {author} {\bibfnamefont {P.}~\bibnamefont {Eden}},
  \bibinfo {author} {\bibfnamefont {C.}~\bibnamefont {Friberg}}, \bibinfo
  {author} {\bibfnamefont {L.}~\bibnamefont {Lonnblad}}, \bibinfo {author}
  {\bibfnamefont {G.}~\bibnamefont {Miu}}, \bibinfo {author} {\bibfnamefont
  {S.}~\bibnamefont {Mrenna}}, \ and\ \bibinfo {author} {\bibfnamefont
  {E.}~\bibnamefont {Norrbin}},\ }\href {\doibase
  10.1016/S0010-4655(00)00236-8} {\bibfield  {journal} {\bibinfo  {journal}
  {Comput. Phys. Commun.}\ }\textbf {\bibinfo {volume} {135}},\ \bibinfo
  {pages} {238} (\bibinfo {year} {2001})},\ \Eprint
  {http://arxiv.org/abs/hep-ph/0010017} {arXiv:hep-ph/0010017} \BibitemShut
  {NoStop}%
\bibitem [{\citenamefont {Frixione}\ \emph {et~al.}(1993)\citenamefont
  {Frixione}, \citenamefont {Mangano}, \citenamefont {Nason},\ and\
  \citenamefont {Ridolfi}}]{Frixione:1993yw}%
  \BibitemOpen
  \bibfield  {author} {\bibinfo {author} {\bibfnamefont {S.}~\bibnamefont
  {Frixione}}, \bibinfo {author} {\bibfnamefont {M.~L.}\ \bibnamefont
  {Mangano}}, \bibinfo {author} {\bibfnamefont {P.}~\bibnamefont {Nason}}, \
  and\ \bibinfo {author} {\bibfnamefont {G.}~\bibnamefont {Ridolfi}},\ }\href
  {\doibase 10.1016/0370-2693(93)90823-Z} {\bibfield  {journal} {\bibinfo
  {journal} {Phys. Lett. B}\ }\textbf {\bibinfo {volume} {319}},\ \bibinfo
  {pages} {339} (\bibinfo {year} {1993})},\ \Eprint
  {http://arxiv.org/abs/hep-ph/9310350} {arXiv:hep-ph/9310350} \BibitemShut
  {NoStop}%
\bibitem [{\citenamefont {Pumplin}\ \emph {et~al.}(2002)\citenamefont
  {Pumplin}, \citenamefont {Stump}, \citenamefont {Huston}, \citenamefont
  {Lai}, \citenamefont {Nadolsky},\ and\ \citenamefont
  {Tung}}]{Pumplin:2002vw}%
  \BibitemOpen
  \bibfield  {author} {\bibinfo {author} {\bibfnamefont {J.}~\bibnamefont
  {Pumplin}}, \bibinfo {author} {\bibfnamefont {D.~R.}\ \bibnamefont {Stump}},
  \bibinfo {author} {\bibfnamefont {J.}~\bibnamefont {Huston}}, \bibinfo
  {author} {\bibfnamefont {H.~L.}\ \bibnamefont {Lai}}, \bibinfo {author}
  {\bibfnamefont {P.~M.}\ \bibnamefont {Nadolsky}}, \ and\ \bibinfo {author}
  {\bibfnamefont {W.~K.}\ \bibnamefont {Tung}},\ }\href {\doibase
  10.1088/1126-6708/2002/07/012} {\bibfield  {journal} {\bibinfo  {journal}
  {JHEP}\ }\textbf {\bibinfo {volume} {07}},\ \bibinfo {pages} {012} (\bibinfo
  {year} {2002})},\ \Eprint {http://arxiv.org/abs/hep-ph/0201195}
  {arXiv:hep-ph/0201195} \BibitemShut {NoStop}%
\bibitem [{\citenamefont {Gluck}\ \emph {et~al.}(1992)\citenamefont {Gluck},
  \citenamefont {Reya},\ and\ \citenamefont {Vogt}}]{Gluck:1991jc}%
  \BibitemOpen
  \bibfield  {author} {\bibinfo {author} {\bibfnamefont {M.}~\bibnamefont
  {Gluck}}, \bibinfo {author} {\bibfnamefont {E.}~\bibnamefont {Reya}}, \ and\
  \bibinfo {author} {\bibfnamefont {A.}~\bibnamefont {Vogt}},\ }\href {\doibase
  10.1103/PhysRevD.46.1973} {\bibfield  {journal} {\bibinfo  {journal} {Phys.
  Rev. D}\ }\textbf {\bibinfo {volume} {46}},\ \bibinfo {pages} {1973}
  (\bibinfo {year} {1992})}\BibitemShut {NoStop}%
\bibitem [{\citenamefont {Mangano}\ \emph {et~al.}(2003)\citenamefont
  {Mangano}, \citenamefont {Moretti}, \citenamefont {Piccinini}, \citenamefont
  {Pittau},\ and\ \citenamefont {Polosa}}]{Mangano:2002ea}%
  \BibitemOpen
  \bibfield  {author} {\bibinfo {author} {\bibfnamefont {M.~L.}\ \bibnamefont
  {Mangano}}, \bibinfo {author} {\bibfnamefont {M.}~\bibnamefont {Moretti}},
  \bibinfo {author} {\bibfnamefont {F.}~\bibnamefont {Piccinini}}, \bibinfo
  {author} {\bibfnamefont {R.}~\bibnamefont {Pittau}}, \ and\ \bibinfo {author}
  {\bibfnamefont {A.~D.}\ \bibnamefont {Polosa}},\ }\href {\doibase
  10.1088/1126-6708/2003/07/001} {\bibfield  {journal} {\bibinfo  {journal}
  {JHEP}\ }\textbf {\bibinfo {volume} {07}},\ \bibinfo {pages} {001} (\bibinfo
  {year} {2003})},\ \Eprint {http://arxiv.org/abs/hep-ph/0206293}
  {arXiv:hep-ph/0206293} \BibitemShut {NoStop}%
\bibitem [{\citenamefont {Frixione}\ and\ \citenamefont
  {Ridolfi}(1997)}]{Frixione:1997ks}%
  \BibitemOpen
  \bibfield  {author} {\bibinfo {author} {\bibfnamefont {S.}~\bibnamefont
  {Frixione}}\ and\ \bibinfo {author} {\bibfnamefont {G.}~\bibnamefont
  {Ridolfi}},\ }\href {\doibase 10.1016/S0550-3213(97)00575-0} {\bibfield
  {journal} {\bibinfo  {journal} {Nucl. Phys. B}\ }\textbf {\bibinfo {volume}
  {507}},\ \bibinfo {pages} {315} (\bibinfo {year} {1997})},\ \Eprint
  {http://arxiv.org/abs/hep-ph/9707345} {arXiv:hep-ph/9707345} \BibitemShut
  {NoStop}%
\bibitem [{\citenamefont {Chekanov}\ \emph {et~al.}(2007)\citenamefont
  {Chekanov} \emph {et~al.}}]{ZEUS:2007njl}%
  \BibitemOpen
  \bibfield  {author} {\bibinfo {author} {\bibfnamefont {S.}~\bibnamefont
  {Chekanov}} \emph {et~al.} (\bibinfo {collaboration} {ZEUS}),\ }\href
  {\doibase 10.1103/PhysRevD.76.072011} {\bibfield  {journal} {\bibinfo
  {journal} {Phys. Rev. D}\ }\textbf {\bibinfo {volume} {76}},\ \bibinfo
  {pages} {072011} (\bibinfo {year} {2007})},\ \Eprint
  {http://arxiv.org/abs/0706.3809} {arXiv:0706.3809 [hep-ex]} \BibitemShut
  {NoStop}%
\bibitem [{\citenamefont {Aktas}\ \emph {et~al.}(2006)\citenamefont {Aktas}
  \emph {et~al.}}]{H1:2006rre}%
  \BibitemOpen
  \bibfield  {author} {\bibinfo {author} {\bibfnamefont {A.}~\bibnamefont
  {Aktas}} \emph {et~al.} (\bibinfo {collaboration} {H1}),\ }\href {\doibase
  10.1016/j.physletb.2006.05.060} {\bibfield  {journal} {\bibinfo  {journal}
  {Phys. Lett. B}\ }\textbf {\bibinfo {volume} {639}},\ \bibinfo {pages} {21}
  (\bibinfo {year} {2006})},\ \Eprint {http://arxiv.org/abs/hep-ex/0603014}
  {arXiv:hep-ex/0603014} \BibitemShut {NoStop}%
\bibitem [{\citenamefont {Klasen}\ and\ \citenamefont
  {Kramer}(1997)}]{Klasen:1996it}%
  \BibitemOpen
  \bibfield  {author} {\bibinfo {author} {\bibfnamefont {M.}~\bibnamefont
  {Klasen}}\ and\ \bibinfo {author} {\bibfnamefont {G.}~\bibnamefont
  {Kramer}},\ }\href {\doibase 10.1007/s002880050528} {\bibfield  {journal}
  {\bibinfo  {journal} {Z. Phys. C}\ }\textbf {\bibinfo {volume} {76}},\
  \bibinfo {pages} {67} (\bibinfo {year} {1997})},\ \Eprint
  {http://arxiv.org/abs/hep-ph/9611450} {arXiv:hep-ph/9611450} \BibitemShut
  {NoStop}%
\bibitem [{\citenamefont {Klasen}\ \emph {et~al.}(1998)\citenamefont {Klasen},
  \citenamefont {Kleinwort},\ and\ \citenamefont {Kramer}}]{Klasen:1997br}%
  \BibitemOpen
  \bibfield  {author} {\bibinfo {author} {\bibfnamefont {M.}~\bibnamefont
  {Klasen}}, \bibinfo {author} {\bibfnamefont {T.}~\bibnamefont {Kleinwort}}, \
  and\ \bibinfo {author} {\bibfnamefont {G.}~\bibnamefont {Kramer}},\ }\href
  {\doibase 10.1007/s101059800001} {\bibfield  {journal} {\bibinfo  {journal}
  {Eur. Phys. J. direct}\ }\textbf {\bibinfo {volume} {1}},\ \bibinfo {pages}
  {1} (\bibinfo {year} {1998})},\ \Eprint {http://arxiv.org/abs/hep-ph/9712256}
  {arXiv:hep-ph/9712256} \BibitemShut {NoStop}%
\bibitem [{\citenamefont {Klasen}\ \emph {et~al.}(2014)\citenamefont {Klasen},
  \citenamefont {Kramer},\ and\ \citenamefont {Michael}}]{Klasen:2013cba}%
  \BibitemOpen
  \bibfield  {author} {\bibinfo {author} {\bibfnamefont {M.}~\bibnamefont
  {Klasen}}, \bibinfo {author} {\bibfnamefont {G.}~\bibnamefont {Kramer}}, \
  and\ \bibinfo {author} {\bibfnamefont {M.}~\bibnamefont {Michael}},\ }\href
  {\doibase 10.1103/PhysRevD.89.074032} {\bibfield  {journal} {\bibinfo
  {journal} {Phys. Rev. D}\ }\textbf {\bibinfo {volume} {89}},\ \bibinfo
  {pages} {074032} (\bibinfo {year} {2014})},\ \Eprint
  {http://arxiv.org/abs/1310.1724} {arXiv:1310.1724 [hep-ph]} \BibitemShut
  {NoStop}%
\bibitem [{\citenamefont {Badger}\ \emph {et~al.}(2013)\citenamefont {Badger},
  \citenamefont {Biedermann}, \citenamefont {Uwer},\ and\ \citenamefont
  {Yundin}}]{Badger:2012pf}%
  \BibitemOpen
  \bibfield  {author} {\bibinfo {author} {\bibfnamefont {S.}~\bibnamefont
  {Badger}}, \bibinfo {author} {\bibfnamefont {B.}~\bibnamefont {Biedermann}},
  \bibinfo {author} {\bibfnamefont {P.}~\bibnamefont {Uwer}}, \ and\ \bibinfo
  {author} {\bibfnamefont {V.}~\bibnamefont {Yundin}},\ }\href {\doibase
  10.1016/j.physletb.2012.11.029} {\bibfield  {journal} {\bibinfo  {journal}
  {Phys. Lett. B}\ }\textbf {\bibinfo {volume} {718}},\ \bibinfo {pages} {965}
  (\bibinfo {year} {2013})},\ \Eprint {http://arxiv.org/abs/1209.0098}
  {arXiv:1209.0098 [hep-ph]} \BibitemShut {NoStop}%
\bibitem [{\citenamefont {Bern}\ \emph {et~al.}(2012)\citenamefont {Bern},
  \citenamefont {Diana}, \citenamefont {Dixon}, \citenamefont {Febres~Cordero},
  \citenamefont {Hoeche}, \citenamefont {Kosower}, \citenamefont {Ita},
  \citenamefont {Maitre},\ and\ \citenamefont {Ozeren}}]{Bern:2011ep}%
  \BibitemOpen
  \bibfield  {author} {\bibinfo {author} {\bibfnamefont {Z.}~\bibnamefont
  {Bern}}, \bibinfo {author} {\bibfnamefont {G.}~\bibnamefont {Diana}},
  \bibinfo {author} {\bibfnamefont {L.~J.}\ \bibnamefont {Dixon}}, \bibinfo
  {author} {\bibfnamefont {F.}~\bibnamefont {Febres~Cordero}}, \bibinfo
  {author} {\bibfnamefont {S.}~\bibnamefont {Hoeche}}, \bibinfo {author}
  {\bibfnamefont {D.~A.}\ \bibnamefont {Kosower}}, \bibinfo {author}
  {\bibfnamefont {H.}~\bibnamefont {Ita}}, \bibinfo {author} {\bibfnamefont
  {D.}~\bibnamefont {Maitre}}, \ and\ \bibinfo {author} {\bibfnamefont
  {K.}~\bibnamefont {Ozeren}},\ }\href {\doibase
  10.1103/PhysRevLett.109.042001} {\bibfield  {journal} {\bibinfo  {journal}
  {Phys. Rev. Lett.}\ }\textbf {\bibinfo {volume} {109}},\ \bibinfo {pages}
  {042001} (\bibinfo {year} {2012})},\ \Eprint {http://arxiv.org/abs/1112.3940}
  {arXiv:1112.3940 [hep-ph]} \BibitemShut {NoStop}%
\bibitem [{\citenamefont {Abdul~Khalek}\ \emph {et~al.}(2021)\citenamefont
  {Abdul~Khalek} \emph {et~al.}}]{AbdulKhalek:2021gbh}%
  \BibitemOpen
  \bibfield  {author} {\bibinfo {author} {\bibfnamefont {R.}~\bibnamefont
  {Abdul~Khalek}} \emph {et~al.},\ }\href@noop {} {\  (\bibinfo {year}
  {2021})},\ \Eprint {http://arxiv.org/abs/2103.05419} {arXiv:2103.05419
  [physics.ins-det]} \BibitemShut {NoStop}%
\end{thebibliography}%

\end{document}